\documentclass[amsmath,amssymb,aps,prd,onecolumn,groupedaddress]{revtex4-2}
\usepackage{bm}
\usepackage{braket}
\usepackage{graphicx}
\usepackage{mathtools}
\usepackage{dcolumn}

\begin{document}

\title{Manifestations of non-zero Majorana CP violating phases in oscillations of supernova neutrinos}

\author{Artem Popov}
\email{ar.popov@physics.msu.ru}
\affiliation{Department of Theoretical Physics, \ Moscow State University, 119991 Moscow, Russia}
\author{Alexander Studenikin}
\email{studenik@srd.sinp.msu.ru}
\affiliation{Department of Theoretical Physics, \ Moscow State University, 119991 Moscow, Russia}
\affiliation{Joint Institute for Nuclear Research, 141980 Dubna, Russia}
\date{\today}
\begin{abstract}
We investigate effects of non-zero Dirac and Majorana CP violating phases on neutrino-antineutrino oscillations in a magnetic field of astrophysical environments. It is shown that in the presence of strong magnetic fields and dense matter, non-zero CP phases can induce new resonances in the oscillations channels $\nu_e \leftrightarrow \bar{\nu}_e$, $\nu_e \leftrightarrow \bar{\nu}_\mu$ and $\nu_e \leftrightarrow \bar{\nu}_{\tau}$. We also consider all other possible oscillation channels with $\nu_\mu$ and $\nu_\tau$ in the initial state. The resonances can potentially lead to significant phenomena in neutrino oscillations accessible for observation in experiments. In particular, we show that neutrino-antineutrino oscillations combined with Majorana-type CP violation can affect the $\bar{\nu}_e$/$\nu_e$ ratio for neutrinos coming from the supernovae explosion. This effect is more prominent for the normal neutrino mass ordering. The detection of supernovae neutrino fluxes in the future experiments, such as JUNO, DUNE and Hyper-Kamiokande, can give an insight into the nature of CP violation and, consequently, provides a tool for distinguishing the Dirac or Majorana nature of neutrinos.
\end{abstract}
\maketitle

\section{Introduction}\label{sec:1}
CP symmetry implies that the equations of motion of a system remain invariant under the CP-transformation, that is a combination of charge conjugation (C) and parity inversion  (P). In 1964, with the discovery of the neutral kaon decay \cite{Christenson:1964fg}, it was confirmed that CP is not an underlying symmetry of the electroweak interactions theory, thus opening a vast field of research in CP violation.
Currently, CP violation is a topic of intense studies in particle physics that also has important implications in cosmology.
In 1967, Sakharov proved that the existence of CP violation is a necessary condition for generation of the baryon asymmetry through baryogenesis in the early Universe \cite{Sakharov:1967dj}. A review of possible baryogenesis scenarios can be found in \cite{Trodden:1998ym}.

Today we have solid understanding of CP violation in the quark sector, that appears due to the complex phase in the Cabibbo–Kobayashi–Maskawa matrix parametrisation. Its magnitude is expressed by the Jarlskog invariant $\mathcal{J}_{CKM} = (3.18\pm0.15)\times10^{-5}$ \cite{Zyla:2020zbs}, which seems to be excessively small to engender baryogenesis at the electroweak phase transition scale \cite{Trodden:1998ym}. However, in addition to experimentally confirmed CP violation in the quark sector, CP violation in the lepton (neutrino) sector hypothetically exists (see \cite{Branco:2011zb} for a review). Leptonic CP violation is extremely difficult to observe due to weakness of neutrino interactions. In 2019, a first breakthrough happened when T2K collaboration reported constraints on the Dirac CP violating phase in neutrino oscillations \cite{Abe:2019vii}. Hopefully, future gigantic neutrino experiments, such as JUNO, DUNE and Hyper-Kamiokande, will have a good chance significantly improve this results. Note that leptonic CP violation plays an important role in baryogenesis through leptogenesis scenarios \cite{Davidson:2008bu}.

The CP violation pattern in the neutrino sector depends on whether neutrino is a Dirac or  Majorana particle.
The Pontecorvo–Maki–Nakagawa–Sakata neutrino mixing matrix in the most common parameterization has  the following form
	\begin{equation}\label{PMNS}
	U=\begin{pmatrix}
	1 & 0 & 0 \\
	0 & c_{23} & s_{23} \\
	0 & -s_{23} & c_{23}
	\end{pmatrix}
	\begin{pmatrix}
	c_{13} & 0 & s_{13} e^{-i\delta} \\
	0 & 1 & 0 \\
	-s_{13} e^{i\delta} & 0 & c_{13}
	\end{pmatrix}
	\begin{pmatrix}
	c_{12} & s_{12} & 0 \\
	-s_{12} & c_{12} & 0 \\
	0 & 0 & 1
	\end{pmatrix}
	\begin{pmatrix}
	e^{i\alpha_1} & 0 & 0 \\
	0 & e^{i\alpha_2} & 0 \\
	0 & 0 & 1
	\end{pmatrix},
	\end{equation}
where $\delta$ is the Dirac CP violating phase, the additional phases $\alpha_1$ and $\alpha_2$ are the so-called Majorana CP violating phases, which can be non-zero only for the case of Majorana neutrinos, and $c_{ik} = \cos \theta_{ik}$ and $s_{ik} = \sin \theta_{ik}$.
As it was shown in \cite{Bilenky:1980cx,Langacker:1986jv}, it is impossible to observe the Majorana CP phases in the neutrino flavour oscillations (see also \cite{Adhikari:2009ja,Giunti:2010ec} for a recent discussion). Nevertheless, in \cite{Xing:2013woa} the authors stated that in principle it is possible to measure Majorana CP phases in neutrino-antineutrino oscillations due to tiny effects of non-zero neutrino masses. However, the probability of this oscillation process for Majorana neutrinos is extremely low to be observed in the near future. Recall that presently the experiments on neutrinoless double beta-decay are considered to be the most promising potential way of measuring the Majorana CP violating phases.

Thus, to measure the Majorana CP violating phases is a challenging task for future physics. In this paper, we study neutrino-antineutrino oscillations engendered by the interaction with a magnetic field in astrophysical environments and search for possible manifestations of the Majorana CP violating phases in neutrino fluxes from supernovae. We show that the effects of the non-zero Majorana CP violating phases indeed can be observed in the neutrino oscillations. Therefore, the detection of supernovae neutrino fluxes can give an insight into the nature of CP violation and, consequently, provide a tool for distinguishing the Dirac or Majorana nature of neutrinos.

The paper is organized as follows.  In Sec. \ref{sec:2} we present a brief introduction into the theory of Majorana neutrinos interactions and mixing. In Sec. \ref{sec:3} we develop the formalism for description of neutrino oscillations in external fields. Sec. \ref{sec:4} presents our numerical results on effects of the Majorana CP violating phases in neutrino-antineutrino oscillations in astrophysical media. Finally, Sec. \ref{sec:5} summarises our results.

\section{Interactions of Majorana neutrinos}\label{sec:2}

In this section we briefly introduce the theory of Majorana neutrino and its interactions.
An arbitrary spinor can be represented as the sum of two independent chiral components
\begin{equation}
\Psi_D = \Psi_L + \Psi_R
\end{equation}
and has four degrees of freedom (corresponding to particle and antiparticle with two helicities).
The Majorana theory of fermions implies that the left and right components of a field are no longer independent and satisfy the relation: $\Psi_R = \Psi_L^c$. Thus, a Majorana spinor has the following form
\begin{equation}
\Psi_M = \Psi_L + \Psi_L^c
\end{equation}
and the neutrality relation $\Psi_M^c = \Psi_M$ holds. A Majorana field has only two degrees of freedom.

The Majorana mass term is introduced as
\begin{equation}
m_i \overline{\nu_i}\nu_i = m_i \overline{(\nu^L_i)^c}\nu^L_i + m_i \overline{\nu^L_i}(\nu^L_i)^c,
\end{equation}
where $i=1,2,3$. The Majorana-type neutrino mass term can be generated by the seesaw mechanism, which naturally appears in the low-energy limit of certain beyond Standard Model theories (see \cite{Cai:2017mow} for a review). An interesting feature of the Majorana mass term is that it violates the total lepton number by two units, which makes possible lepton number violating processes, such as the neutrinoless double beta-decay (0$\nu\beta\beta$). At the moment, experiments on 0$\nu\beta\beta$ are considered to be the most prospective way to probe the nature of the neutrino mass. Furthermore, these experiments are potentially sensitive to the Majorana CP violating phases difference $\alpha_1-\alpha_2$ \cite{Branco:2011zb}.

The Majorana condition $\nu^c_i=\nu_i$ puts significant constraints on the structure of the flavour neutrino fields. Unlike in the case of Dirac neutrinos, when the mixing matrix for the right component of the field can be introduced arbitrarily since right-handed neutrinos are sterile, for Majorana neutrinos the following relations hold
\begin{eqnarray}
\nu_{\alpha}^L &=& \sum_{i} U_{\alpha i} \nu_i^L,\\
\nu_{\alpha}^R &=& (\nu_{\alpha}^L)^c =  \sum_{i} U_{\alpha i}^* (\nu_i^L)^c,
\end{eqnarray}
and
\begin{equation}
\nu_{\alpha} = \nu_{\alpha}^L + \nu_{\alpha}^R = \sum_{i} U_{\alpha i} \nu_i^L + \sum_{i} U_{\alpha i}^* (\nu_i^L)^c,
\end{equation}
where $\alpha = e,\mu,\tau$. Obviously, the flavour fields satisfy the Majorana condition $\nu^c_{\alpha}=\nu_{\alpha}$.

Now consider the Majorana neutrino interactions. It is known that massive neutrinos must possess a non-zero anomalous magnetic moment, and therefore interact with a magnetic field (see \cite{Giunti:2014ixa} for a review on the electromagnetic properties of neutrinos). The Majorana neutrino interaction with a magnetic field has the following form
\begin{equation}\label{mag_field_int}
\mathcal{L}_{mag} = -\sum_{i,k} \mu_{ik}\left[ \overline{(\nu_i^L)^c} \bm{\Sigma}\bm{B} \nu_k^L + \overline{\nu_i^L} \bm{\Sigma}\bm{B} (\nu_k^L)^c \right] =
-\sum_{\alpha,\beta} \mu_{\alpha \beta}\left[ \overline{(\nu_{\alpha}^L)^c} \bm{\Sigma}\bm{B} \nu_{\beta}^L + \overline{\nu_{\alpha}^L} \bm{\Sigma}\bm{B} (\nu_{\beta}^L)^c \right].
\end{equation}
It is clear from the form of the Lagrangian that the interaction with a magnetic field can induce neutrino-antineutrino oscillations $\nu_{\alpha} \leftrightarrow \bar{\nu}_{\beta}$. The Majorana condition imposes certain  constraints on neutrino magnetic moments. Since a Majorana neutrino is a truly neutral particle, it cannot posses diagonal electric and magnetic dipole form factors. However, non-diagonal entries are possible, in particular the transition magnetic moments. The magnetic moments matrix of a Majorana neutrino $\mu_{ik}$ is antisymmetric and Hermitian, and then has only non-diagonal entries which are purely imaginary quantities: $\mu_{ik} = i|\mu_{ik}| = -\mu_{ki}$ for $i\neq k$. Numerical values of the neutrino magnetic moments are discussed in Sec. \ref{sec:4}. For a thorough review of neutrino electromagnetic properties and spin oscillations see \cite{Giunti:2014ixa} and references therein.

Interactions of neutrinos with matter are described by the Lagrangian
\begin{equation}
\mathcal{L}_{mat} = -\sum_{\alpha} V^{(f)}_{\alpha} \overline{\nu_{\alpha}}\gamma_{0}(1+\gamma_5)\nu_{\alpha},
\end{equation}
where
\begin{equation}
V^{(f)} = \operatorname{diag} \left(\frac{G_F n_e}{\sqrt{2}} - \frac{G_F n_n}{2\sqrt{2}}, - \frac{G_F n_n}{2\sqrt{2}}, - \frac{G_F n_n}{2\sqrt{2}} \right)
\end{equation}
is the Wolfenstein potential. Here we consider a normal electrically neutral matter composed of electrons, protons and neutrons.
Since for a Majorana field the relation $\bar{\Psi}\gamma_{\mu}\Psi = 0$ holds, one can replace $(1+\gamma_5)$ with $\gamma_5$. Then, the matter interaction Lagrangian takes the following forms for Majorana and Dirac neutrino respectively:
\begin{equation}\label{matter_int_majorana}
\mathcal{L}_{mat}^M = -\sum_{\alpha} V^{(f)}_{\alpha} \left[\overline{\nu^L_{\alpha}} \gamma_0 \nu_{\alpha}^L - \overline{(\nu^L_{\alpha})^c} \gamma_0 (\nu_{\alpha}^L)^c  \right],
\end{equation}
\begin{equation}\label{matter_int_dirac}
\mathcal{L}_{mat}^D = -\sum_{\alpha} V^{(f)}_{\alpha} \overline{\nu^L_{\alpha}} \gamma_0 \nu_{\alpha}^L.
\end{equation}


It is well known \cite{Bilenky:1980cx,Langacker:1986jv} that it is not possible to distinguish between Dirac and Majorana neutrinos in studies of the neutrino flavour oscillations. Our studies below are initiated by an expectation that the neutrino-antineutrino oscillations induced by a magnetic field could provide an appropriate setup to probe the nature of the neutrino mass term.
In particular, we show that under certain realistic astrophysical conditions the Majorana CP violating phases affect neutrino oscillations pattern.

The effects of non-zero Majorana CP violating phases in neutrino-antineutrino oscillations have been studied before in \cite{Xing:2013woa}. The authors considered neutrino-antineutrino oscillations in vacuum induced by the Majorana mass term $m_{\alpha\beta}\overline{\nu^c}_{\alpha} \nu_{\beta}$. Despite the fact that the probabilities of such oscillations are strongly suppressed by the factor of order $m^2/E^2$, they still can be possibly used to determine the magnitudes of both Dirac and Majorana CP violating phases in future terrestrial experiments, provided that neutrino is a Majorana fermion. In \cite{Balantekin:2007es,Gava:2008rp} effects of non-zero Dirac CP violating phase were studied without accounting for the interaction with a magnetic field. It was shown that the magnitude of the Dirac CP phase does not affect supernovae neutrino fluxes unless muon and tau (anti)neutrino fluxes differ at the neutrinosphere.

In turn, here below we show that supernovae neutrino fluxes can carry significant information about CP violation, given that neutrino magnetic moments are large enough to enable substantial $\nu \leftrightarrow \bar{\nu}$ oscillations in the supernova envelope.
In what follows we  develop a consistent approach to the description of neutrino spin (or neutrino-antineutrino in the Majorana case) oscillations in astrophysical environments in the three neutrino framework.

\section{Formalism}\label{sec:3}
In this section we extend the formalism developed in \cite{Popov:2019nkr} to account for the transition magnetic moments and neutrino interactions with matter. We derive the following system of Dirac equations for the massive neutrino states
\begin{eqnarray}\label{dirac_equation}
(i\gamma^{\mu} \partial_{\mu} - m_i - V^{(m)}_{ii} \gamma^{0}\gamma_5)\nu_i(x) -\sum_{k \neq i} \big(\mu_{ik}\bm{\Sigma}\bm{B} + V^{(m)}_{ik}  \gamma^{0}\gamma_5\big)\nu_k(x) = 0,
\end{eqnarray}
where $V^{(m)} = U^{\dag}V^{(f)}U$ is the matter potential in the mass basis, $i,k = \{1,2,3\}$. In the presence of electron matter component $n_e$ and/or interaction of the transition magnetic moments with a magnetic field these three equations are coupled. As a result,  the neutrino mass states under these conditions are non-stationarity.
Eq. (\ref{dirac_equation}) can be rewritten in the Hamiltonian form:
\begin{equation}\label{equation_hamiltonian}
i\frac{\partial}{\partial t} \nu(x) = \begin{pmatrix}
H_{11} & H_{12} & H_{13} \\
H_{21} & H_{22} & H_{23} \\
H_{31} & H_{32} & H_{33}
\end{pmatrix} \nu(x) = H \nu(x),
\end{equation}
where $\nu = (\nu_e, \nu_{\mu}, \nu_{\tau})^T$ and
\begin{equation}
H_{ik} = \delta_{ik} \gamma_0 \bm{\gamma}\bm{p} + m_i\delta_{ik} \gamma_0 + \mu_{ik}\gamma_0\bm{\Sigma}\bm{B} + V_{ik}^{(m)}\gamma_5.
\end{equation}

Our goal is to calculate the probabilities of oscillations between different neutrino and antineutrino states. The probabilities are expressed as
\begin{eqnarray}\label{prob_1}
P(\nu_{\alpha}^s \rightarrow \nu_{\beta}^{s'}) = \big|\braket{\nu_{\beta}^{s'}(0)|\nu_{\alpha}^s(x)}\big|^2 = \Big|  \sum_{i,k} \left(U^{s'}_{\beta k}\right)^{*} U_{\alpha i}^s \braket{\nu_k^{s'}(0)|\nu_i^s(x)}\Big|^2,
\end{eqnarray}
where $U^s$ are the mixing matrices for left ($s=L$) and right ($s=R$) neutrinos. Note that in the Majorana case right neutrino is an antineutrino: $\nu_i^R = (\nu_i^L)^c$ and then $U^L = U$, $U^R = U^*$.

Now we focus on the calculation of the amplitudes $\braket{\nu_k^{s'}(0)|\nu_i^s(x)}$ using Eq. (\ref{dirac_equation}).
Consider the neutrino mass states with a definite helicity
\begin{eqnarray} \nonumber
\ket{\psi^L_1(0)} &=& \begin{pmatrix}
\ket{L} \\
0 \\
0
\end{pmatrix}, \ \ \
\ket{\psi^R_1(0)} = \begin{pmatrix}
\ket{R} \\
0 \\
0
\end{pmatrix},
\\ \label{initial_conditions}
\ket{\psi^L_2(0)} &=& \begin{pmatrix}
0 \\
\ket{L} \\
0
\end{pmatrix}, \ \ \
\ket{\psi^R_2(0)} = \begin{pmatrix}
0 \\
\ket{R} \\
0
\end{pmatrix},
\\ \nonumber
\ket{\psi^L_3(0)} &=& \begin{pmatrix}
0 \\
0 \\
\ket{L}
\end{pmatrix}, \ \ \
\ket{\psi^R_3(0)} = \begin{pmatrix}
0 \\
0 \\
\ket{R}
\end{pmatrix},
\end{eqnarray}
where $\ket{L}$ and $\ket{R}$ are the eigenvectors of the helicity operator $\bm{\Sigma}\bm{p}/p$, the eigenvalues are $-1$ and $+1$, respectively. We consider astrophysical neutrinos with energies of order 10 MeV  \cite{Mirizzi:2015eza} and mass to be  1.1 eV, that is  equal to the upper-bound reported by the KATRIN collaboration \cite{Aker:2019uuj}. Thus, ultrarelativistic assumption is justified, and we can write out
\begin{equation}
\ket{L} = \frac{1}{\sqrt{2}} \begin{pmatrix}
0 \\
-1 \\
0 \\
1
\end{pmatrix}, \ \ \
\ket{R} = \frac{1}{\sqrt{2}} \begin{pmatrix}
1 \\
0 \\
1 \\
0
\end{pmatrix}.
\end{equation}

The formal solution of the evolution equation with the initial states (\ref{initial_conditions}) is

\begin{equation}
\ket{\psi^L_i(x)} = e^{-i H x}\ket{\psi^L_i(0)}.
\end{equation}
Indices $i$ and $L$ refer only to the initial conditions (\ref{initial_conditions}), and, since the massive neutrino state with a definite helicity is generally not a stationary quantum state, for $x>0$ the state $\ket{\psi_i^L(x)}$ is, strictly speaking, no longer a massive neutrino state with a certain polarization. This state rather accounts for possible transitions between neutrino mass states and helicity states due to interactions with matter and the magnetic field. The amplitudes of the transitions in Eq. (\ref{prob_1}) are
\begin{equation}\label{amplitudes}
\braket{\nu^{s'}_k(0)|\nu_i^s(x)} =  \braket{\psi_k^{s'}(0)|\psi_i^s(x)}.
\end{equation}

The easiest way to compute the amplitudes of interest (\ref{amplitudes}) is to use the eigendecomposition of the Hamiltonian
\begin{equation}
H =\sum_n E_n \ket{n}\bra{n}, \ \ \ \ H \ket{n}= E_n\ket{n}
\end{equation}
and the following relation for the matrix exponential
\begin{equation}
e^{-i H x} = \sum_n e^{-i E_n x}\ket{n}\bra{n} = \sum_n e^{-i E_n x} P_n,
\end{equation}
where
\begin{equation}
P_n = \ket{n}\bra{n}.
\end{equation}
The amplitudes of the transitions between massive neutrino states with a definite helicity can be represented in the form of the plane wave decomposition
\begin{eqnarray}
\braket{\nu_k^{s'}(0)|\nu_i^s(x)} = \sum_n \braket{\psi_k^{s'}(0)|P_n|\psi_i^s(0)} e^{-i E_n x} = \sum_n C_{nki}^{ss'} e^{-i E_n x},
\end{eqnarray}
where the coefficients
\begin{equation}
C_{nki}^{ss'} = \braket{\psi_k^{s'}(0)|P_n|\psi_i^s(0)}
\end{equation}
are used.

The probabilities of neutrino oscillations are
\begin{equation}\label{prob}
P(\nu_\alpha^s \rightarrow \nu_\beta^{s'}) = \Big| \sum_n \sum_{i,k} \left(U_{\beta k}^{s'}\right)^* U_{\alpha i}^s C_{nki}^{s s'} e^{-i E_n x} \Big|^2.
\end{equation}
They can be expressed in the explicit form
\begin{widetext}
\begin{equation}\label{prob_real}
P(\nu_{\alpha}^s \rightarrow \nu_{\beta}^{s'};x) = \delta_{\alpha \beta}\delta_{s s'} - 4 \sum_{n>m} \operatorname{Re}(A_{\alpha\beta nm}^{s s'}) \sin^2\left(\frac{\pi x}{ L^{osc}_{nm} }\right)
+ 2 \sum_{n>m} \operatorname{Im}(A_{\alpha\beta nm}^{s s'})  \sin\left(\frac{2\pi x} {L^{osc}_{nm}}\right),
\end{equation}
\end{widetext}
where
\begin{equation}\nonumber
A_{\alpha\beta nm}^{s s'} = \sum_{i,j,k,l} \left(U_{\beta k}^{s'}\right)^* U_{\alpha i}^s U_{\beta l}^{s'} \left(U_{\alpha j}^s\right)^* \big(C_{nki}^{ss'}\big)^*C_{mlj}^{s s'}
\end{equation}
and $L^{osc}_{nm} = 2\pi/ (E_n - E_m)$.

This formula generalizes the well-known expression for the probabilities of vacuum neutrino oscillations. A similar expression was obtained in \cite{Lichkunov:2020} for the case of Dirac neutrinos with only diagonal magnetic moments. The last term of (\ref{prob_real}) is T violating and, provided that CPT symmetry is conserved, is also CP violating. CPT conservation, however, is not true for the case of the neutrino interaction with particle-antiparticle asymmetric media, such as a supernova environment, because of the matter-induced (the extrinsic) CPT violation \cite{Jacobson:2003wc}. Thus, for a realistic astrophysical environment we can not assume that the last term of (\ref{prob_real}) encapsulates the CP violating effects.

For the further considerations it is useful to introduce two additional quantities. Firstly, using (\ref{prob}) one can calculate the amplitudes of oscillations:
\begin{equation}\label{amplitude}
P(\nu_{\alpha}^s \rightarrow \nu_{\beta}^{s'})_{max} = \Big( \sum_n |\mathcal{I}_{n\alpha\beta}^{ss'}| \Big)^2,
\end{equation}
	where
\begin{equation}
\mathcal{I}_{n\alpha\beta}^{s s'} = \sum_{i,k} (U_{\beta k}^{s'})^* U^s_{\alpha i} C^{s s'}_{nki}.
\end{equation}
Secondly, from (\ref{prob_real}) we derive the distance-averaged probabilities:
\begin{equation}\label{prob_avg}
\langle P(\nu_{\alpha}^s \rightarrow \nu_{\beta}^{s'}) \rangle = \delta_{\alpha \beta}\delta_{s s'} - 2 \sum_{n>m} \operatorname{Re}(A_{\alpha\beta nm}^{s s'}).
\end{equation}

In what follows we apply the developed formalism for the Majorana neutrino oscillations to study neutrino fluxes in astrophysical media, peculiar, for instance, for supernovae. Note that the developed approach with a straightforward  modification can be also applied to the case of Dirac neutrinos.

\section{Astrophysical applications}\label{sec:4}

Here below we present the numerical results on neutrino-antineutrino oscillations in supernovae neutrino fluxes.
First we analyse possible resonances in the neutrino-antineutrino oscillations channels. Then we investigate the effects of these resonances on the flavour composition of the neutrino fluxes for a certain supernova model.

The magnitudes of the oscillations parameters in the PMNS matrix (\ref{PMNS}) are given in Table \ref{tab:table1} (see \cite{Zyla:2020zbs,Esteban:2018azc}).
\begin{table}[h]
\caption{\label{tab:table1}%
Neutrino oscillation parameters according to \cite{Esteban:2018azc}.}
\begin{ruledtabular}
	\begin{tabular}{c|c|c|c|c|c}
		\textrm{Parameter}&
		\textrm{$\sin^2 \theta_{12}$}&
		\textrm{$\sin^2 \theta_{23}$}&
		\textrm{$\sin^2 \theta_{13}$}&
		\textrm{$\Delta m^2_{12}  / \text{eV} ^2$}&
		\textrm{$\left| \Delta m^2_{13} \right|$/eV$^2$}\\
		\colrule
		Value & 0.310 & 0.558 & 0.022 & 7.39$\times 10^{-5}$ & 2.52$\times 10^{-3}$
	\end{tabular}
\end{ruledtabular}
\end{table}

The magnetic moments matrix in the case of Majorana neutrinos is antisymmetric and consists of the purely imaginary entries (see [14] for a detailed discussion). In the studies below we use the following representation:
\begin{equation}
\mu_{ij}=\begin{pmatrix}
0 & i|\mu_{12}| & i|\mu_{13}| \\
-i|\mu_{12}| & 0 & i|\mu_{23}| \\
-i|\mu_{13}| & -i|\mu_{23}| & 0
\end{pmatrix}.
\end{equation}

The best terrestrial experiment upper bounds on the neutrino magnetic moments, obtained by the GEMMA reactor neutrino experiment \cite{Beda:2012zz} and  Borexino collaboration \cite{Borexino:2017fbd} by measuring the solar neutrino fluxes, are on the level $\mu_{\nu} < 2.8 \div 2.9 \times 10^{-11} \mu_{B}$. An order of magnitude more stringent upper bound is provided by the observed properties of the globular cluster stars \cite{Raffelt:1990pj,Viaux:2013hca,Arceo-Diaz:2015pva}. For our further analyses we fix the values of the transition magnetic moments in the neutrino mass basis accordingly: $|\mu_{12}|=|\mu_{13}|=|\mu_{23}| = 10^{-12} \mu_B$. The particular features of the neutrino oscillations described below are generally appropriate for the case of an arbitrary choice of non-zero transition magnetic moments.

\subsection{Resonances in neutrino-antineutrino oscillations}
Consider the amplitudes (\ref{amplitude}) of neutrino-antineutrino oscillations. Since the neutrino magnetic moments are, generally speaking, small, we are principally interested in oscillations under the extreme external conditions peculiar to astrophysical objects, supernovae in particular. A supernova inner region is characterized by the baryon number density $n_B$ that is of order  $10^{30}$ cm$^{-3}$ and even higher, with $n_B = n_p + n_n$, where $n_p$ and $n_n$ are the proton and neutron number densities, respectively. For a neutral media, as one of a supernova, the proton density $n_p$ is equal to the electron density $n_e$. Magnetic fields during a core-collapse can reach magnitudes up to $10^{15}-10^{16}$ G right after $\approx 9$ ms after bounce (see \cite{Akiyama:2002xn}). In our analysis below we use a more conservative value for the magnetic field $B \sim 10^{13}$ G and also chose the baryon number density $n_B = 10^{32}$ cm$^{-3}$.

We are particularly interested in the neutrino-antineutrino oscillations engendered by a magnetic field because the MSW resonances in the neutrino flavour oscillations are not possible under extreme densities in the inner supernova regions. However, the flavour oscillations dominate in the outer regions, where the densities are lower and the magnetic field is excessively weak to engender the spin-flip transitions. The MSW effect contribution is considered in Subsec. B.

Fig. 1 shows the amplitudes (\ref{amplitude}) of the neutrino-antineutrino oscillations as functions of the electron fraction $Y_e = n_e/n_B$ for the CP violating phases given by  $\delta = 0$, $\alpha_{1} = 0$, $\alpha_{2} = 0$. The resonant curve in Fig. 1 reproduces the well-known resonant behaviour of the spin-flavor ($RSF$) conversion studied in \cite{Akhmedov:1988uk, Lim:1987tk}, with the resonance in the $\nu_e \to \bar{\nu}_{\mu}$ channel for $Y_e \approx 0.5$, described by \cite{Akhmedov:1988uk, Lim:1987tk}
\begin{equation}
P_{max} = \mu^2 B^2/( (\mu B)^2 + \Delta H^2 ),  \ \  \ \Delta H = \sqrt{2} G_F n_B (1 - 2 Y_e) - \Delta m^2 \cos2\theta/2p.
\end{equation}

However, the CP conserving values $\delta = 0$ and $\delta = \pi$ are disfavored at the 95$\%$ confidence level by the T2K collaboration \cite{Abe:2019vii}, and the data shows a preference for near maximal CP violation $\delta = \pi/2$. Therefore, it is worth to proceed with consideration of the CP violating effects in neutrino spin oscillations.

The amplitudes of the neutrino-antineutrino oscillations for the case of non-zero Dirac CP violating phase are shown in Fig. 2. The resonance in the channel $\nu_e \to \bar{\nu}_{\mu}$ is persistent for all values of $\delta$. There is also a new resonance in the channel $\nu_e \to \bar{\nu}_{e}$ that appears at $Y_e \approx 0.35$. The location of the resonance does not depend either on the magnetic field strength $B$ or the baryon density $n_B$ and the neutrino energy $p$. This resonance occurs even for values of the Dirac CP violating phase which are only slightly different from the CP conserving values, i.e. $\delta = 0$ or $\pi$. Thus we can expect significant $\nu_e \rightarrow \bar{\nu}_e$ conversions at a certain point of a supernova if neutrinos are Majorana particles, Dirac CP violating phase $\delta$ is non-zero and the interaction with the stellar magnetic field is strong enough ($B\sim 10^{12} \div 10^{13} \ G$).

Although the effects of the Dirac CP violating phase may be important in astrophysics, the phase itself is likely to be measured in terrestrial experiments (such as JUNO, DUNE and Hyper-Kamiokande) in the near future.

We focus below on manifestations of the Majorana CP violating phases. The amplitudes (\ref{amplitude}) of neutrino-antineutrino oscillations for the cases $\delta = 0$ and $\delta = \pi/2$ correspondingly as functions of the electron fraction for the particular values of Majorana CP violating phases $\alpha_1$ and $\alpha_2$ are shown in Fig. 3 and Fig. 4 . A particular feature is that now the resonant peak in the $\nu_e \leftrightarrow \bar{\nu}_{\tau}$ oscillations can appear at $Y_e \approx 0.5$ for certain magnitudes of the Majorana phases in addition to the resonances in the $\nu_e \leftrightarrow \bar{\nu}_{e}$ and $\nu_e \leftrightarrow \bar{\nu}_{\mu}$ conversions. Fig. 5 and Fig. 6 present the amplitudes of $\nu \leftrightarrow \bar{\nu}$ oscillations as functions of both $\alpha_1$ and $\alpha_2$.

From (\ref{amplitude}) we have also found that the amplitudes of $\nu_e \leftrightarrow \bar{\nu}_{\mu}$ and $\nu_e \leftrightarrow \bar{\nu}_{\tau}$ conversions do not depend on $\delta$. However, the appearance of a non-zero $\delta$ breaks the central symmetry of $P(\nu_e \rightarrow \bar{\nu}_e)_{max}$ shown in Fig. 5.

The resonant values of the electron fraction $Y_e = 0.35$ and $Y_e = 0.5$ found above are robust and do not depend on the magnetic field strength, neutrino energy or baryon number density. However, the width of the resonant curves $P_{max}(Y_e)$ varies with the field-density ratio $\mu B/ G_F n_B$: the resonances become wider as it increases. Additionally, the $\nu_e \leftrightarrow \bar{\nu}_e$ resonance peak is inherently narrower than two others.

It is important to estimate the scale $L_{res}$ of the supernova region where $\nu \leftrightarrow \bar{\nu}$ oscillations exhibit the resonance behavior and compare it to the corresponding oscillations length $L_{osc}$. In the case $L_{osc} \ll L_{res}$ there is no room for the resonance behavior of the $\nu \leftrightarrow \bar{\nu}$ oscillations in the considered astrophysical environment.
The scale of the resonant region is given by
\begin{equation}\label{scale_1}
L_{res} \approx \left(\frac{d Y_e}{dr}\right)^{-1} \Delta Y_e,
\end{equation}
where $\Delta Y_e$ is the width of the resonant curve $P_{max}(Y_e)$. The value $dY_e/dr$ typically is of order $10^{-8}$ cm$^{-1}$ \cite{Mirizzi:2015eza} and it is natural to modify (\ref{scale_1}) accordingly

\begin{equation}\label{scale_2}
L_{res} \approx \left(\frac{d Y_e/dr}{10^{-8} \operatorname{cm}^{-1}} \right)^{-1} \Delta Y_e \times 10^3 \operatorname{km}.
\end{equation}

Our estimations show that $\Delta Y_e \approx 0.3$ for the $\nu_e \leftrightarrow \bar{\nu}_e$ oscillations, and $\Delta Y_e \approx 1$  for the $\nu_e \leftrightarrow \bar{\nu}_{\mu}$ and $\nu_e \leftrightarrow \bar{\nu}_{\tau}$ oscillations for the set of parameters described above (see Figs. 1-4).  The corresponding resonant regions scales $L_{res}$ are found to be 30 km and 100 km, respectively. At the same time, the oscillation length $L_{osc}$ is below 1 km. Thus, the resonance oscillations $\nu\leftrightarrow\bar{\nu}$ are expected to proceed not far from the neutrinosphere.

Finally, we conclude that depending on the particular values of the Dirac and Majorana CP violating phases three resonances may occur: $\nu_e \leftrightarrow \bar{\nu}_{e}$ resonance (at $Y_e = 0.35$), $\nu_e \leftrightarrow \bar{\nu}_{\mu}$ resonance and $\nu_e \leftrightarrow \bar{\nu}_{\tau}$ resonance (both at $Y_e = 0.5$).
Note that the amplitudes of oscillations described above do not significantly vary with the neutrino energy, provided that it is greater than 0.1 MeV. For neutrinos with energies far below this threshold the oscillations patterns become drastically different, but this is not the case of interest for the neutrinos from supernovae. The effects of different mass hierarchies have also been found to be subtle. We qualitatively show below how the appearance of the new resonances affects observable neutrino fluxes.

\subsection{Effects of CP violation on supernova neutrino fluxes}
\label{subsec_B}

Consider the influence of the $\nu \leftrightarrow \bar{\nu}$ oscillations on the neutrino fluxes emitted at the later stages of a supernova evolution. We compute neutrino fluxes as follows

\begin{equation}\label{fluxes}
\Phi_{\nu_{\alpha}} = \sum_{\beta} \Phi_{\nu_\beta}^0 P(\nu_{\beta} \rightarrow \nu_{\alpha}),
\end{equation}
where $\alpha, \beta = e,\bar{e},\mu,\bar{\mu},\tau,\bar{\tau}$ and $\Phi^0_{\nu_\beta}$ are the neutrino fluxes at the neutrinosphere. As it follows from (\ref{scale_2}), the resonant regions scales are much larger then the oscillation length: $L_{res} \gg L_{osc}$. Thus, the probabilities  $P(\nu_{\beta} \rightarrow \nu_{\alpha})$ can be replaced with the averaged probabilities (\ref{prob_avg}). Besides, both the $\nu \to \bar{\nu}$ and MSW resonant regions are smaller than the distance between them. Thus, in the evaluation of the neutrino fluxes one can first calculate the neutrino fluxes within the inner supernova region accounting for the $\nu \leftrightarrow \bar {\nu}$ oscillations, and then proceed with accounting for the MSW oscillations.

For our simulation we use the following values of the initial neutrino fluxes: $\Phi_{\nu_{e}}^0 = \Phi_{\bar{\nu}_e}^0 = 4.1 \times 10^{51}$ erg/sec, $\Phi_{\nu_x}^0 = 7.9 \times 10^{51}$ erg/sec, $\Phi_{\nu_\mu, \nu_\tau, \overline{\nu}_\mu, \overline{\nu}_\tau}^0 = \Phi_{\nu_x}^0$ \cite{Keil:2002in}. Supernovae models predict that $Y_e \approx 0.4$ during the later stages of a supernova explosion \cite{Nunokawa:1997ct}. This value lies between the resonant values $Y_e = 0.35$ and $Y_e = 0.5$. Since the resonance at $Y_e = 0.35$ is substantially narrower than the resonance at $Y_e = 0.5$, we expect that the latter contribution is more significant. Accounting for the probability conservation relation $\sum_{\alpha} P(\nu_{\alpha} \to \nu_{\beta}) = 1$, one can simplify (\ref{fluxes}) as follows

\begin{equation}\label{fluxes_simplified}
\Phi_{\nu_\alpha} = \Phi^0_{\nu_x} + (\Phi^0_{\nu_e} - \Phi^0_{\nu_x}) \left[ P(\nu_e \rightarrow \nu_{\alpha}) + P(\overline{\nu}_e \rightarrow \nu_{\alpha} ) \right].
\end{equation}

Since it is impossible to detect the muon and tau neutrino and antineutrino fluxes separately for the energy range of a supernova neutrino emission (see in \cite{Mirizzi:2015eza}), one have to construct observables using only $\Phi_{\nu_e}$, $\Phi_{\bar{\nu}_e}$ and $\Phi_{\nu_x}$.  From our analysis it follows that the most pronounced effect of the Majorana CP violation phases can appear in the ratio of $\bar{\nu_e}$ and $\nu_e$ fluxes that can be written in the following form

\begin{equation}\label{asymmetry}
\frac{\Phi_{\bar{\nu}_e}}{\Phi_{\nu_e}} = \frac{ 1 + (\Phi^0_{\nu_e}/\Phi^0_{\nu_x} - 1) \langle P(\bar{\nu}_e \to \bar{\nu}_e) \rangle + (\Phi^0_{\nu_e}/\Phi^0_{\nu_x} - 1) \langle P(\nu_e \to \bar{\nu}_e) \rangle }{1 + (\Phi^0_{\nu_e}/\Phi^0_{\nu_x} - 1) \langle P(\nu_e \to \nu_e)\rangle + (\Phi^0_{\nu_e}/\Phi^0_{\nu_x} - 1) \langle P(\bar{\nu}_e \to \nu_e) \rangle }.
\end{equation}

Note that a deviation from unity of the ratio $\Phi_{\bar{\nu}_e}/\Phi_{\nu_e}$ ($\Phi_{\bar{\nu}_e}/\Phi_{\nu_e} \neq 1$) indicates for CP violating effects. Two types of the CP violating effects can take place in neutrino oscillations: the extrinsic (matter-induced) and intrinsic \cite{Jacobson:2003wc}. We are interested only in the intrinsic CP violation, i.e. the effects of non-zero CP violating phases. Our numerical results presented below show that for neutrino energies above 0.1 MeV the extrinsic contribution to (\ref{asymmetry}) is negligible.

Fig. 7 (Left) shows the $\bar{\nu}_e/\nu_e$ ratio, calculated based on (\ref{asymmetry}),  as a function of the Dirac CP phase $\delta$ for the case $\alpha_{1}=0$, $\alpha_{2}=0$. The $\bar{\nu}_e/\nu_e$ ratio scarcely reaches 12\% at $\delta = \pi/2$ and $3\pi/2$. The relative insignificance of the effect follows from the fact that $\nu_e \leftrightarrow \bar{\nu}_e$ oscillations amplitudes are suppressed at $Y_e = 0.4$. Thus, one can neglect the effect of the Dirac CP phase in the evaluation the $\bar{\nu}_e/\nu_e$ ratio.

Consider now  the $\bar{\nu}_e/\nu_e$ ratio as a function of the Majorana CP violating phases for $\delta = 0$. The results are shown in Fig. 7 (Right). Except the regions around $\alpha_2 = \alpha_1 \pm \pi$, the neutrino-antineutrino oscillations induce significant asymmetry between $\nu_e$ and $\bar{\nu}_e$ fluxes, that peaks at 50\%. From  (\ref{fluxes_simplified}) it also follows that the muon and tau neutrino and antineutrino fluxes are still approximately equal to each other after oscillations: $\Phi_{\nu_{\mu}}=\Phi_{\nu_{\tau}}=\Phi_{{\bar\nu}_{\mu}}=\Phi_{{\bar\nu}_{\tau}} $. Thus, neutrino-antineutrino oscillations in a magnetic field of the inner supernova region can indeed induce significant asymmetry between the $\nu_e$ and $\bar{\nu}_e$ fluxes.

The next step is to compute the neutrino fluxes outside the supernova.
For this, one has to account for the Mikheev-Smirnov-Wolfenstein oscillations, which take place in outer supernova regions with relatively low densities. The adiabatic solution for the neutrino fluxes in the discussed case yields  \cite{Mirizzi:2015eza}:

\begin{equation}
\begin{matrix*}[l]
\Phi^{out}_{\nu_e} = \Phi_{\nu_x} & \mbox{(NH)}, \\
\Phi^{out}_{\nu_e} = s_{12}^2	\Phi_{\nu_e} + c_{12}^2 \Phi_{\nu_x} & \mbox{(IH)}, \\
\Phi^{out}_{\bar{\nu}_e} = c_{12}^2	\Phi_{\bar{\nu}_e} + s_{12}^2 \Phi_{\nu_x} & \mbox{(NH)}, \\
\Phi^{out}_{\bar{\nu}_e} = \Phi_{\nu_x} & \mbox{(IH)},
\end{matrix*}
\end{equation}
where NH and IH refer to the normal and inverted mass hierarchy respectively, and $\Phi_{\nu_\mu, \nu_\tau, \overline{\nu}_\mu, \overline{\nu}_\tau} = \Phi_{\nu_x}$.

The final results for $\bar{\nu}_e/\nu_e$ fluxes  ratio in a supernova emission are shown in Fig.8. It follows that the magnetic field induced disproportion in the fluxes (see the above comments to Fig. 7) becomes smeared after the MSW oscillations, specially for the case of the inverted mass hierarchy: the maximal values for the ration are $\Phi^{out}_{\bar{\nu}_e}/\Phi^{out}_{\nu_e}=0.23$ for NH and $\Phi^{out}_{\bar{\nu}_e}/\Phi^{out}_{\nu_e}=0.23$ for IH. Table \ref{table2} shows $\bar{\nu}_e/\nu_e$ ratio characteristic values (the ``No magnetic field'' column stands for the case of $B=0$ and the ``No CP'' column stands for the case of $\alpha_1=0$, $\alpha_2=0$).
The minimal and maximal values of the ratio $\Phi^{out}_{\bar{\nu}_e}/\Phi^{out}_{\nu_e}$ (depending on the Majorana CP phases) are also shown in Table \ref{table2}.

\begin{table}[h]
	\caption{\label{table2}%
		$\bar{\nu}_e/\nu_e$ ratio characteristic values.}
	\begin{ruledtabular}
		\begin{tabular}{c|c|c|c|c}
			\textrm{$\Phi^{out}_{\bar{\nu}_e}/\Phi^{out}_{\nu_e}$}&
			\textrm{No magnetic field}&
			\textrm{Min}&
			\textrm{No CP}&
			\textrm{Max}\\
			\colrule
			NH & 0.67 & 0.64 & 0.74 & 0.87 \\
			IH & 1.18 & 0.9 & 0.97 & 1.0 \\
		\end{tabular}
	\end{ruledtabular}
\end{table}

\section{Conclusion}\label{sec:5}
In this paper we study the Majorana neutrinos oscillations in astrophysical media with an emphasis on the CP violating effects. The semi-analytical expressions for the neutrino oscillations probabilities are obtained. It is shown that the appearance of the non-zero CP violating phases can give rise to new resonances in the neutrino-antineutrino oscillations channels, namely: $\nu_e \leftrightarrow \overline{\nu}_e$ resonance at $Y_e = 0.35$ that appears for both cases of the non-zero Dirac and Majorana phases; and $\nu_e \leftrightarrow \overline{\nu}_{\mu},{\nu}_{\tau}$ resonances at $Y_e = 0.5$ that is possible only for the case of the non-zero Majorana CP phases. These resonant values of $Y_e$ are persistent and do not depend on the magnetic field strength $B$ or neutrino energy $p$.

The results obtained are applied to a particular physical situation: the oscillations of neutrinos during the cooling stage of a supernova explosion. It is shown that neutrino-antineutrino oscillations in the near-neutrinosphere high-density region of a supernova can significantly modify the resulting outcoming neutrino fluxes. Particularly, under certain non-zero values of the Majorana CP phases, the $\bar{\nu}_e/\nu_e$ ratio reaches magnitudes up to 1.5 (as is shown in Fig. 7) within the inner supernova region. After the consequent MSW oscillations, the effect becomes less pronounced, specially for the case of the inverted neutrino mass hierarchy, but still is present. Our results, however, only roughly estimate the $\bar{\nu}_e/\nu_e$ ratio. For a precise calculation one have to account for several more factors, particularly a realistic supernova density and the magnetic field profiles, as well as the collective effects in neutrino oscillations. Latter can be important, since nonlinear feedback due to self-interaction can enhance a small effect. For now, we point out that the quantity $\Phi_{\bar{\nu}_e}/\Phi_{\nu_e}$ can potentially be an important observable for the supernova neutrino experiments.

One of the important new results is the conclusion that observations of the ratio of supernovae fluxes $\Phi_{\bar{\nu}_e}/\Phi_{\nu_e}$ in the future large volume neutrino detectors, such as JUNO and Hyper-Kamiokande, may provide a tool for distinguishing between the Dirac and Majorana nature of neutrinos.

Future neutrino experiments will hopefully not only provide us high-statistics measurements of neutrino fluxes directly from a supernova explosion, but also will be sensitive to the diffuse supernova neutrino background (DSNB). We argue that DNSB flavour composition may depend on the CP violating phases. Although to extract information about CP violation from DNSB is an utterly complicated task, the diffuse neutrinos provide us a continuous source of experimental data, while a supernova explosion is quite a rare event in the Galaxy.

\section*{Acknowledgements}
The work is supported by the Russian Foundation for Basic Research under grant No.
20-52-53022-GFEN-a. The work of A.P. is supported by the Foundation for the Advancement of
Theoretical Physics and Mathematics “BASIS” under grant No. 19-2-6-209-1.

\newpage

\begin{figure}
	\begin{minipage}{0.49\linewidth}
		\includegraphics[width=1\linewidth]{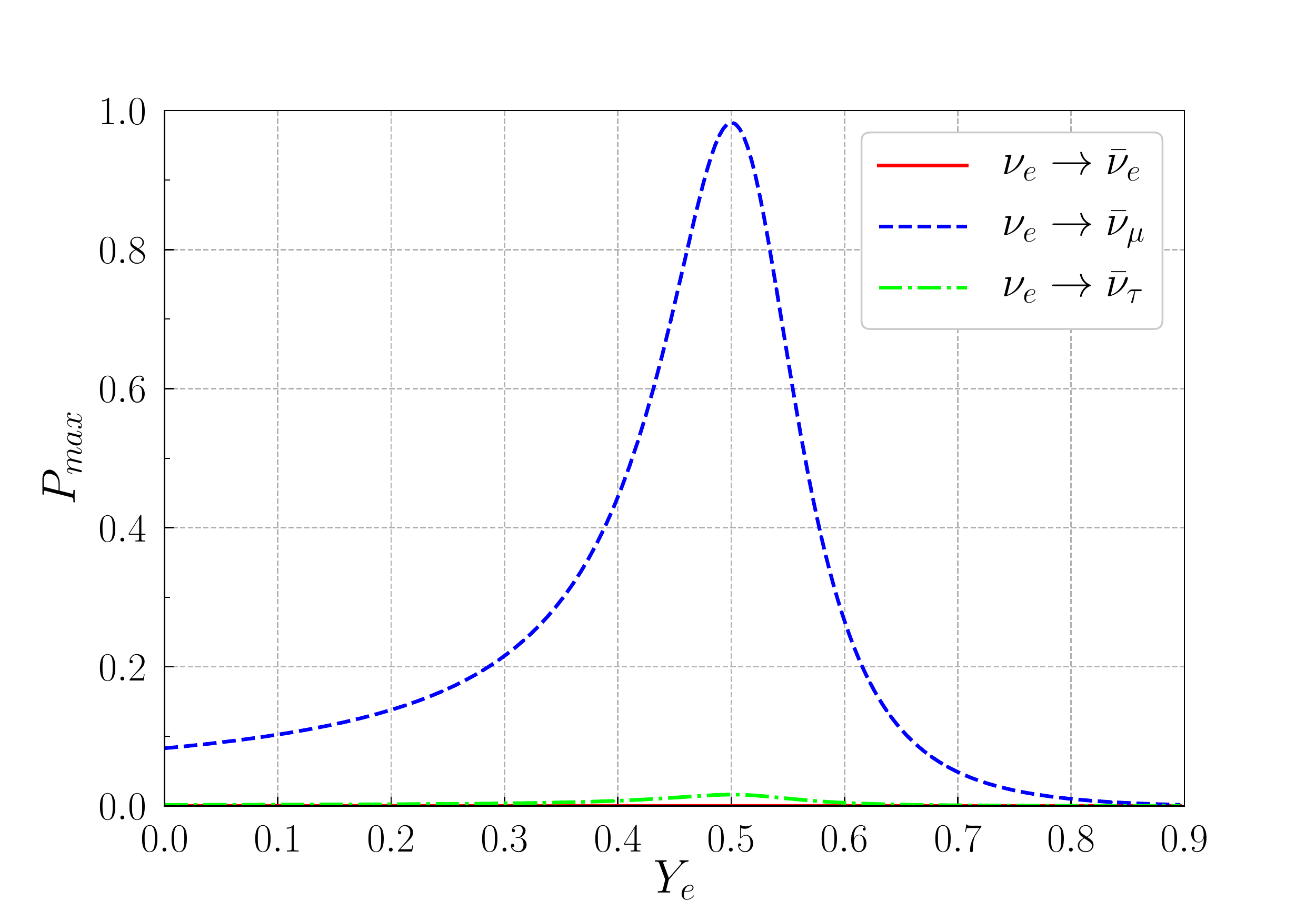}
		\caption{The amplitudes of neutrino-antineutrino oscillations as functions of the electron fraction $Y_e$ for the case of CP conservation.}
	\end{minipage}
\end{figure}

\begin{figure*}
	\begin{minipage}{0.49\linewidth}
		\includegraphics[width=1\linewidth]{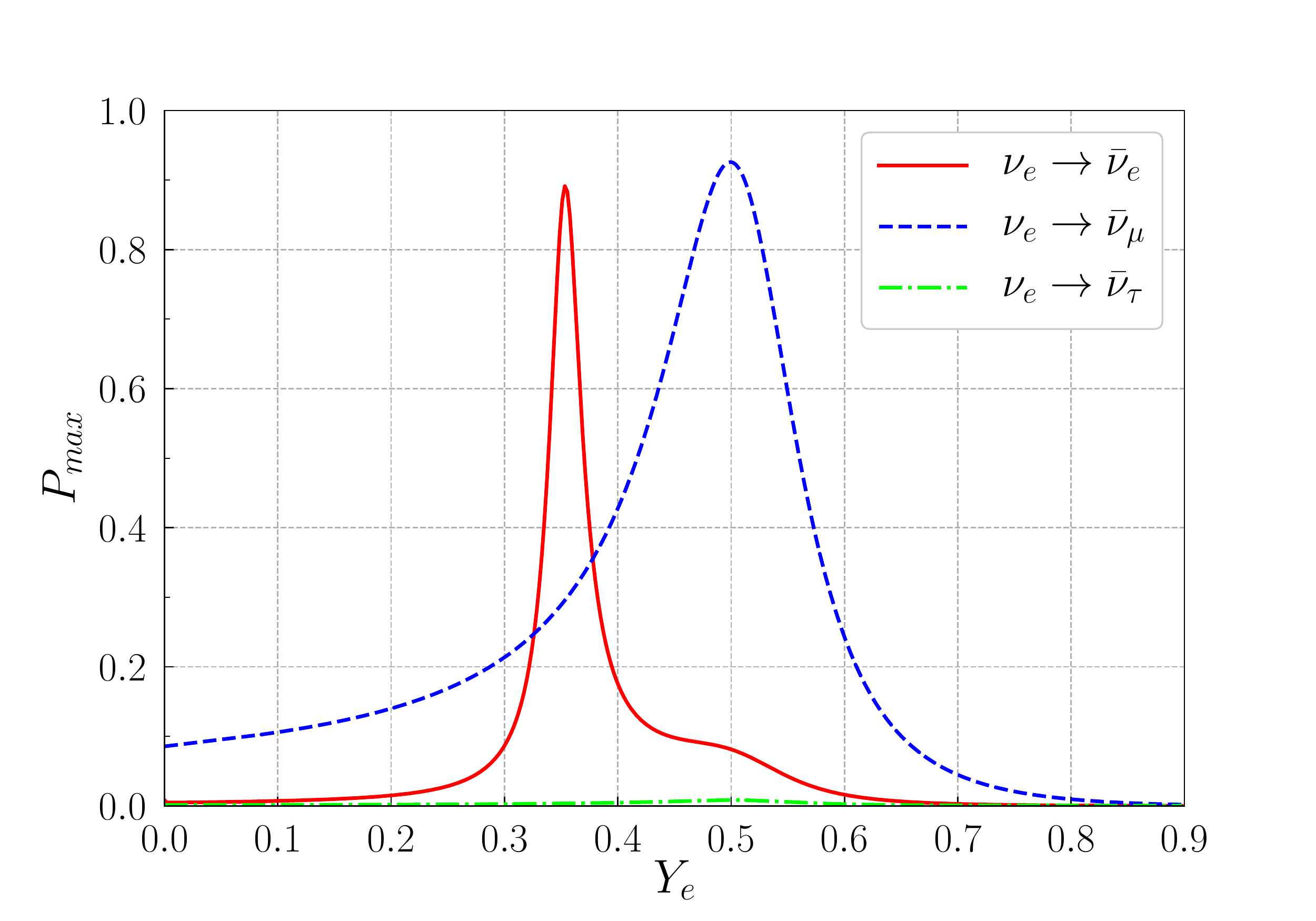}
	\end{minipage}
	\begin{minipage}{0.49\linewidth}
		\includegraphics[width=1\linewidth]{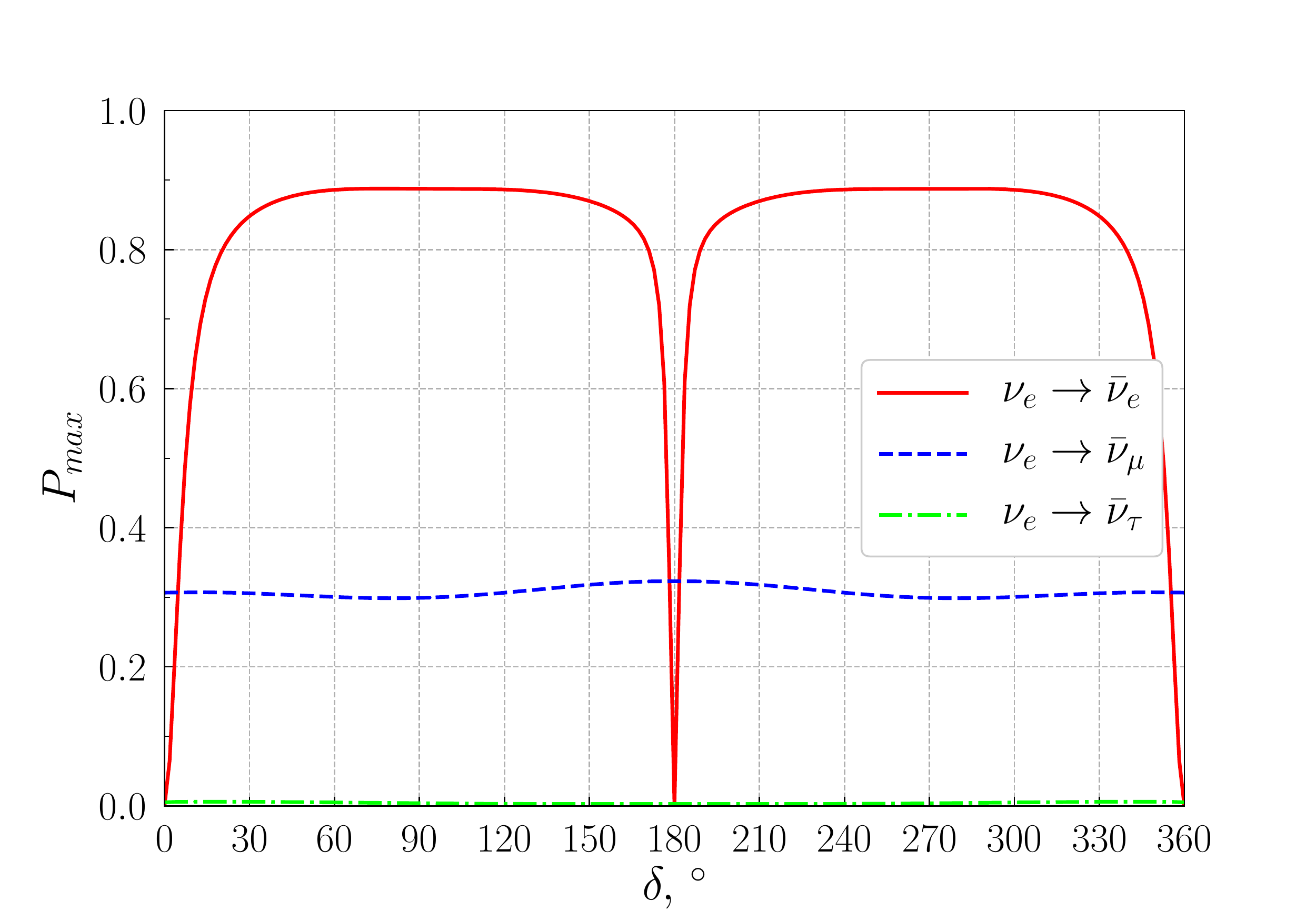}
	\end{minipage}
	\caption{Left: The amplitudes of neutrino-antineutrino oscillations as functions of the electron fraction $Y_e$ for the case of $\delta = \pi/2$, $\alpha_1=0$ and $\alpha_2=0$. Right: The amplitudes of neutrino-antineutrino oscillations as functions of the Dirac CP violating phase $\delta$ for the case $Y_e = 0.35$, $\alpha_1=0$ and $\alpha_2=0$.}
\end{figure*}

\begin{figure*}
	\begin{minipage}{0.49\linewidth}
		\includegraphics[width=1\linewidth]{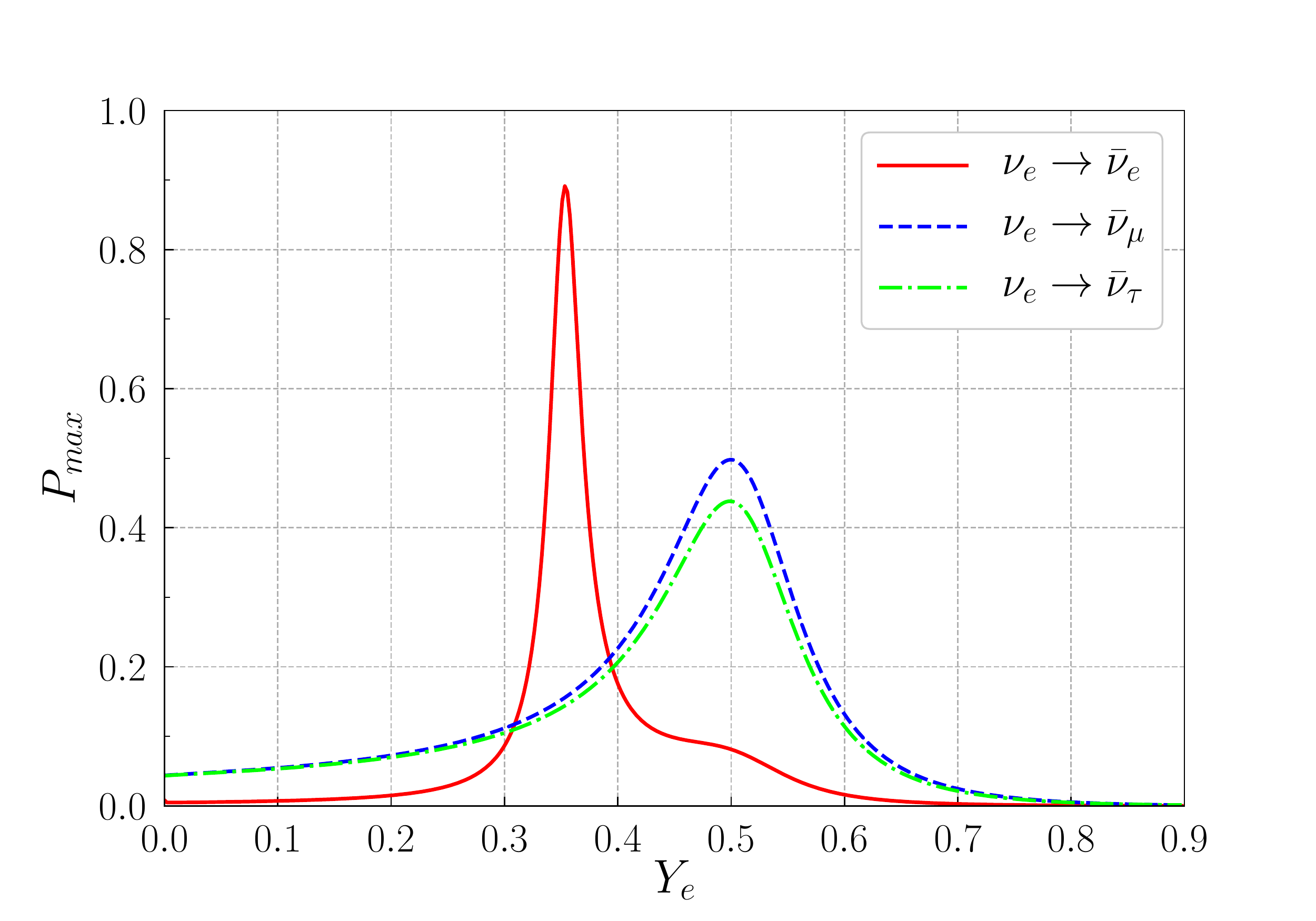}
	\end{minipage}
	\begin{minipage}{0.49\linewidth}
		\includegraphics[width=1\linewidth]{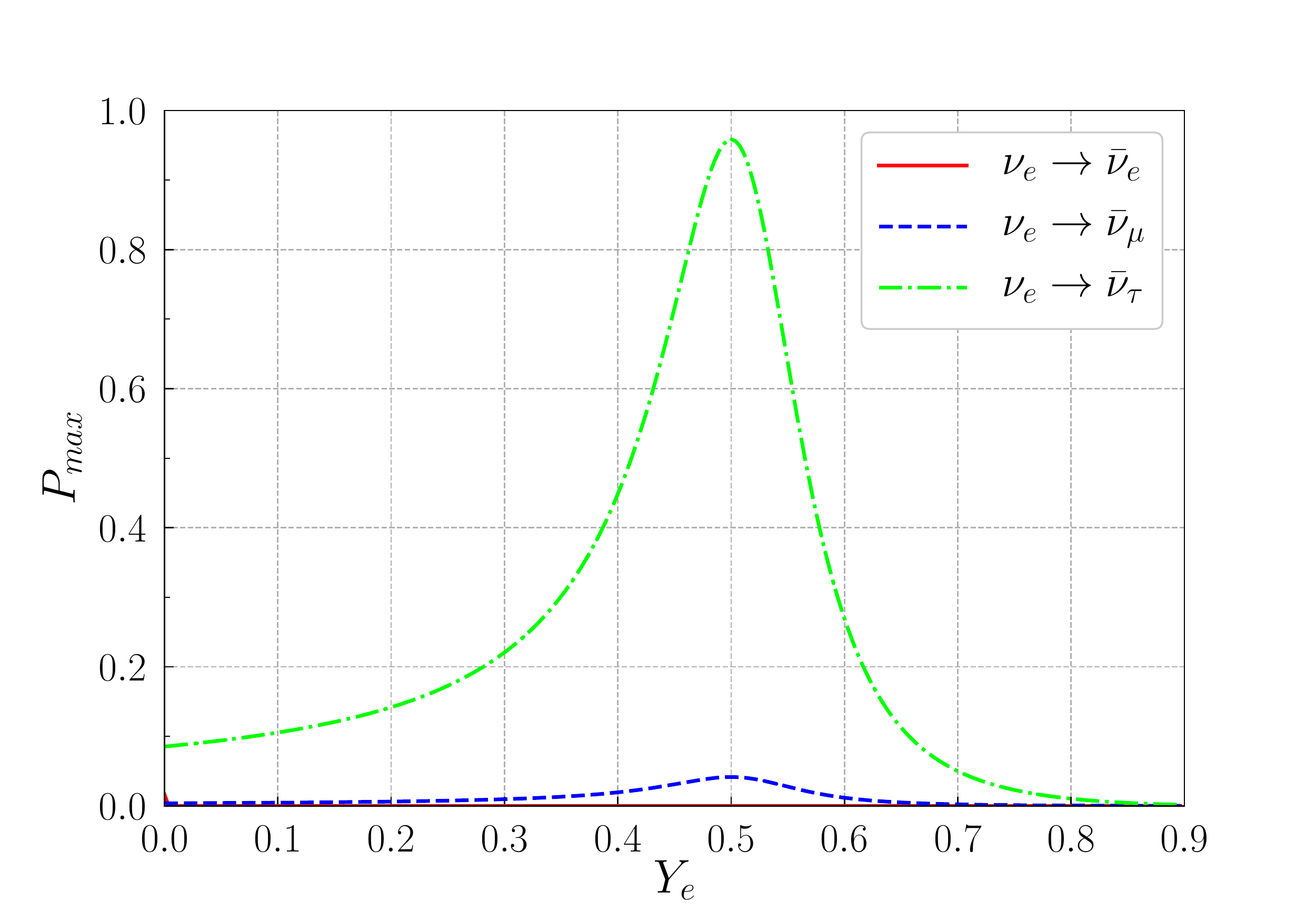}
	\end{minipage}
	\caption{The amplitudes of neutrino-antineutrino oscillations as functions of the electron fraction $Y_e$ for the case of $\delta = 0$. Left: $\alpha_1=\alpha_2=\pi/2$. Right: $\alpha_1=\alpha_2=\pi$.}
\end{figure*}

\begin{figure*}
	\begin{minipage}{0.49\linewidth}
		\includegraphics[width=1\linewidth]{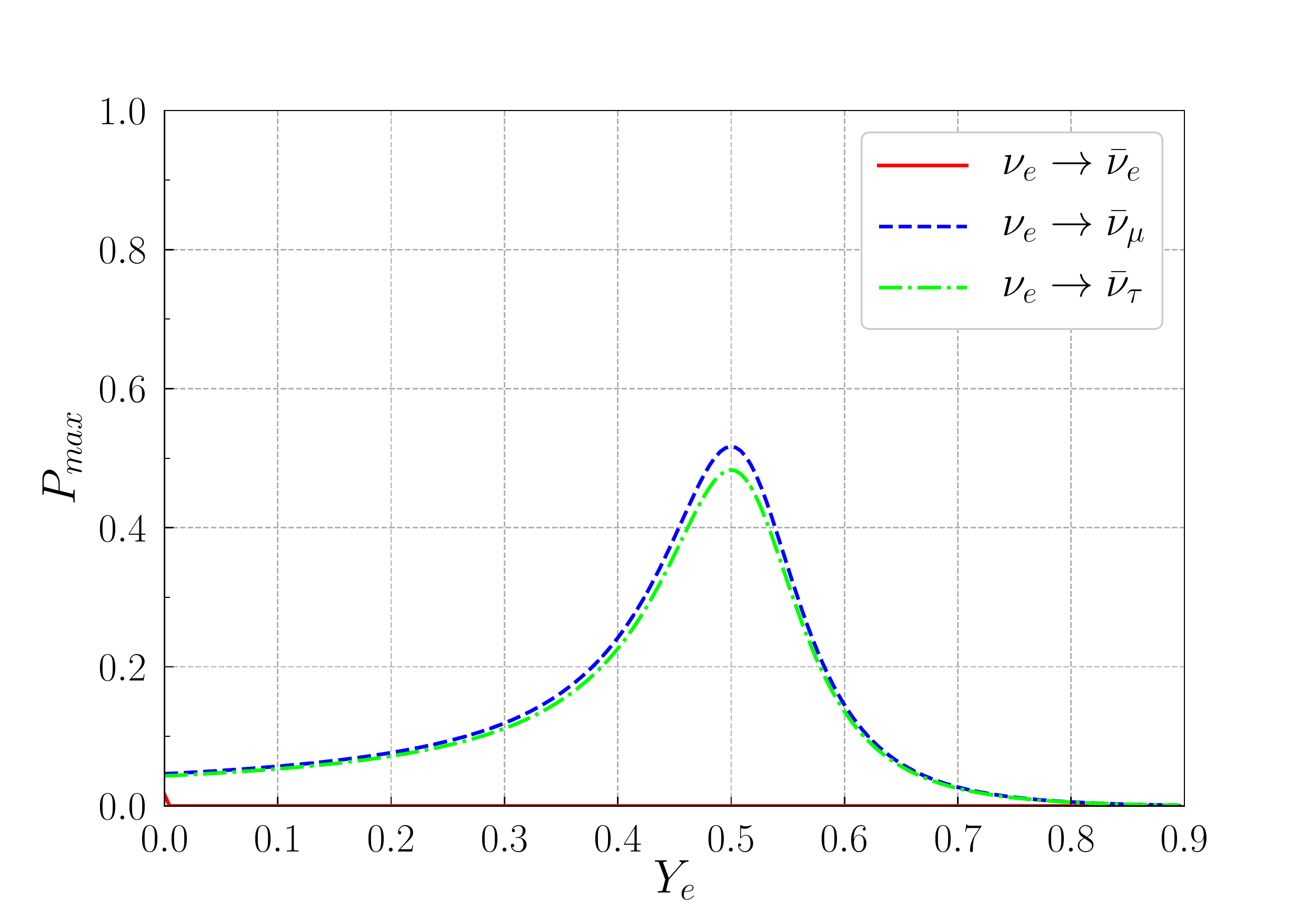}
	\end{minipage}
	\begin{minipage}{0.49\linewidth}
		\includegraphics[width=1\linewidth]{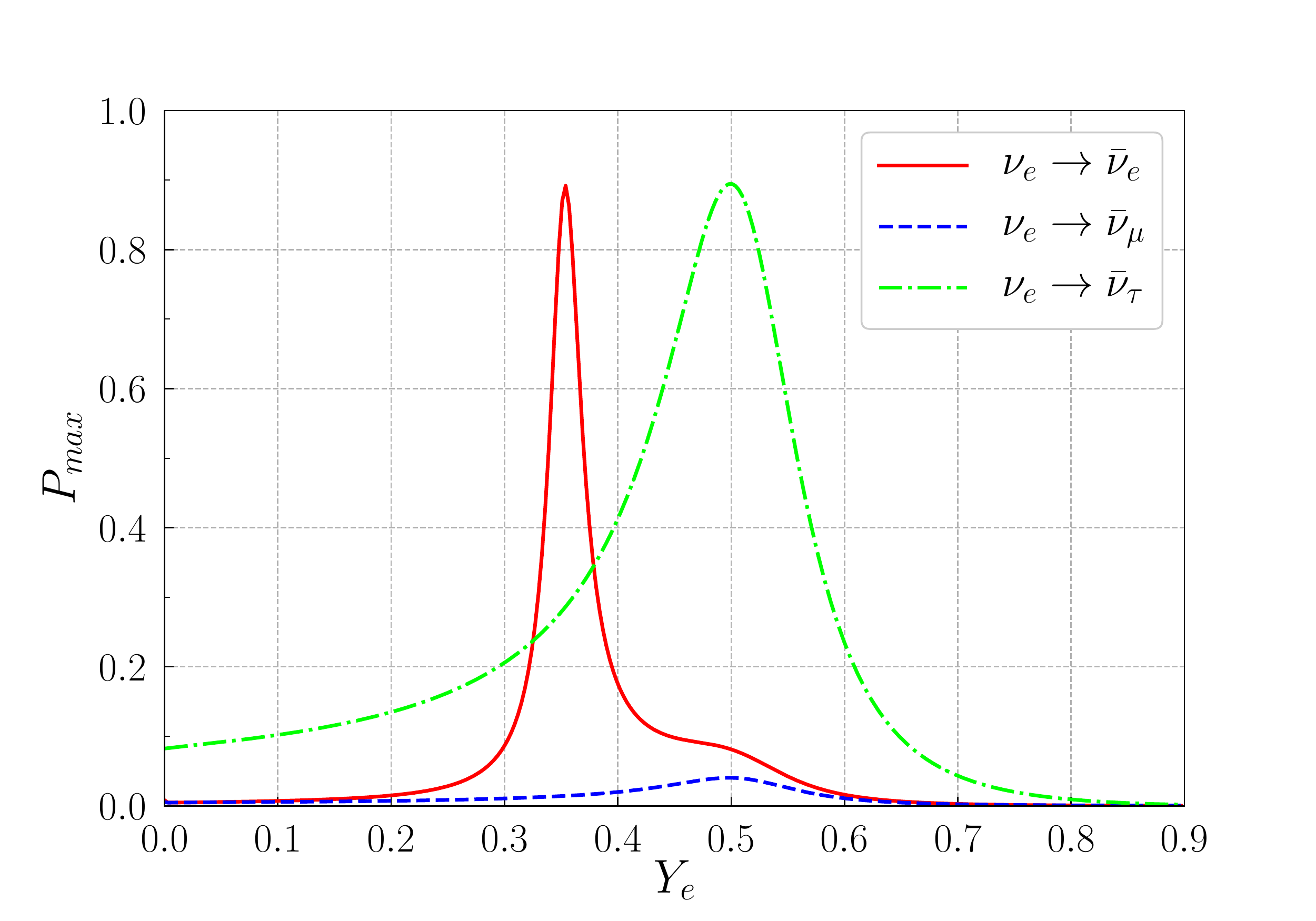}
	\end{minipage}
	\caption{The amplitudes of neutrino-antineutrino oscillations as functions of the electron fraction $Y_e$ for the case of $\delta = \pi/2$. Left: $\alpha_1=\alpha_2=\pi/2$. Right: $\alpha_1=\alpha_2=\pi$.}
\end{figure*}

\begin{figure*}
	\begin{minipage}{0.49\linewidth}
		\includegraphics[width=1\linewidth]{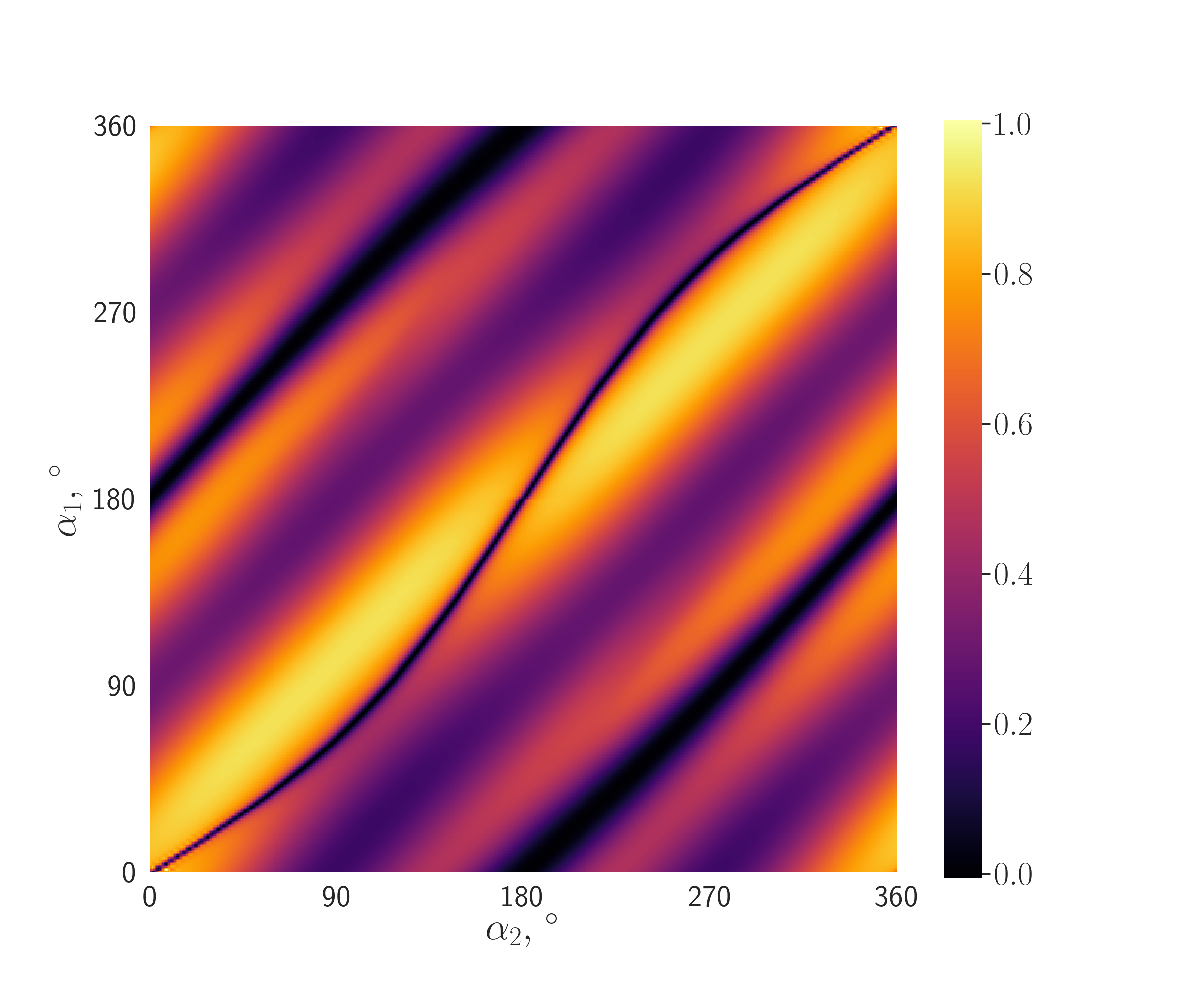}
	\end{minipage}
	\begin{minipage}{0.49\linewidth}
		\includegraphics[width=1\linewidth]{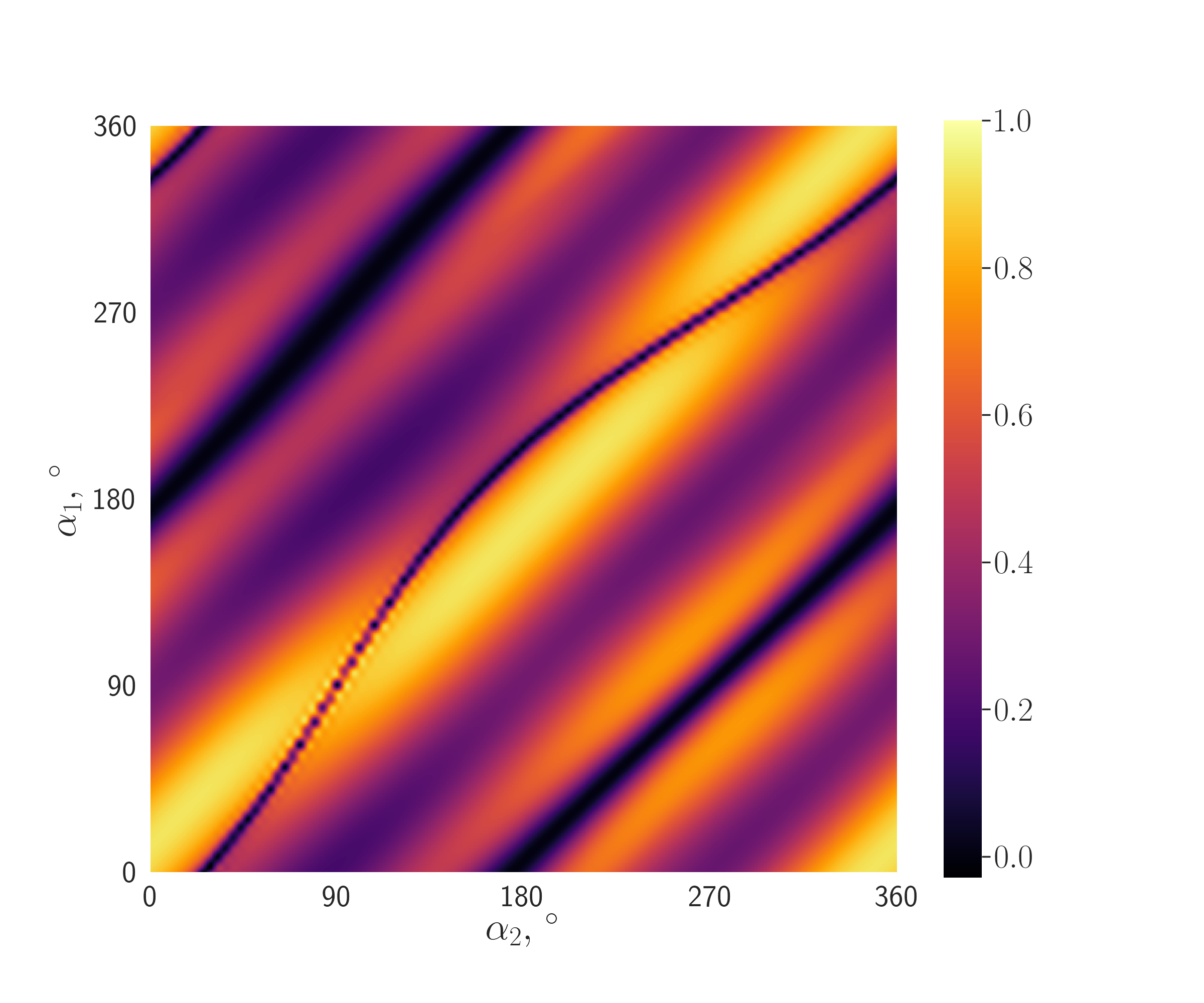}
	\end{minipage}
	\caption{The amplitude of $\nu_e \to \overline{\nu}_e$ oscillations for the case $Y_e = 0.35$ as a function of the Majorana CP violating phases $\alpha_1$ and $\alpha_2$. Left: $\delta = 0$. Right: $\delta = \pi/2$. Black and yellow represent amplitudes equal to 0 and 1, respectively. }
\end{figure*}

\begin{figure*}
	\begin{minipage}{0.49\linewidth}
		\includegraphics[width=1\linewidth]{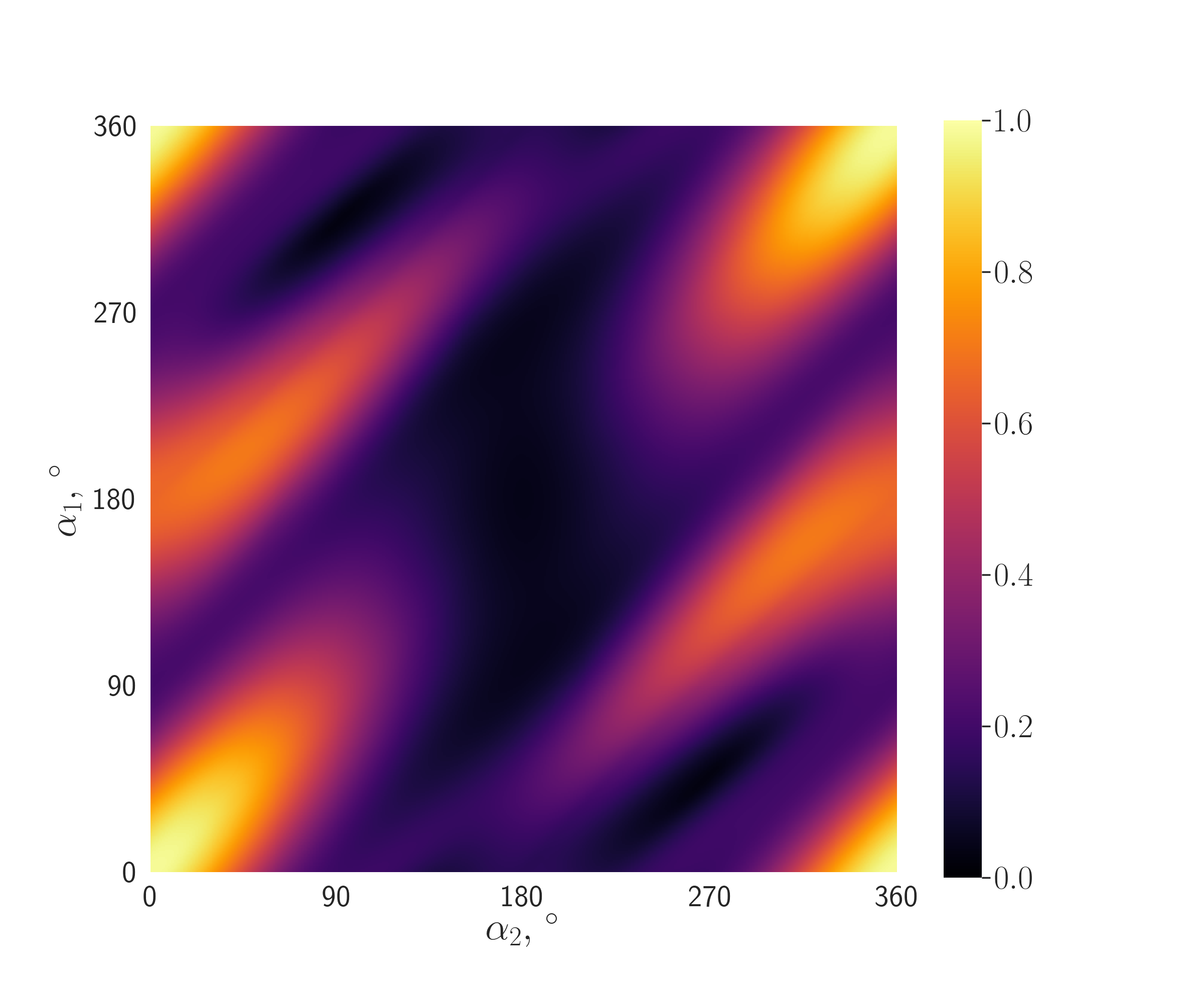}
	\end{minipage}
	\begin{minipage}{0.49\linewidth}
		\includegraphics[width=1\linewidth]{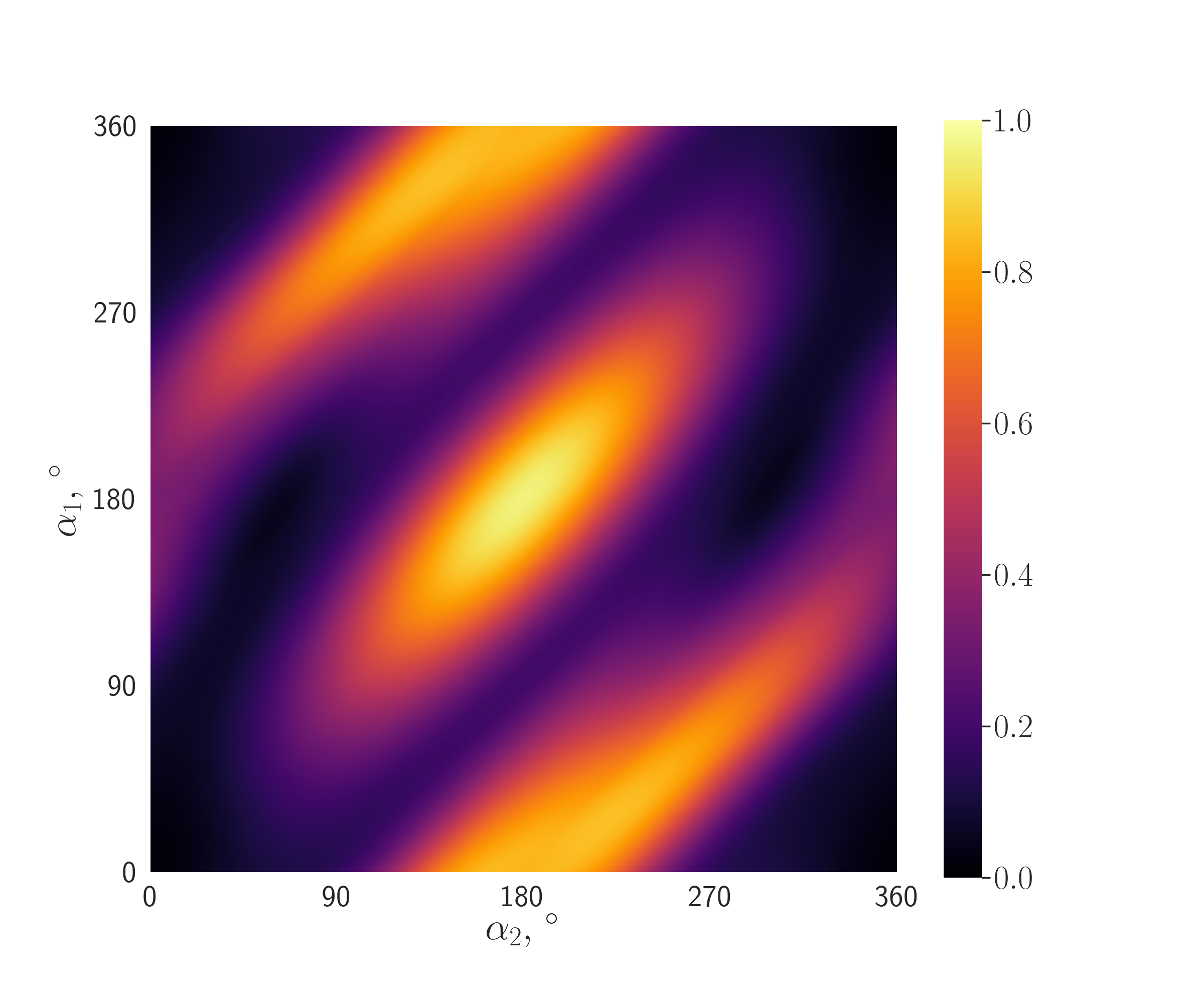}
	\end{minipage}
	\caption{Left: The amplitude of $\nu_e \to \overline{\nu}_{\mu}$ oscillations for the case $Y_e=0.5$ as a function of the Majorana CP violating phases $\alpha_1$ and $\alpha_2$. Right: Same, but for the amplitude of $\nu_e \to \overline{\nu}_{\tau}$ oscillations.}
\end{figure*}

\begin{figure*}
	\begin{minipage}{0.49\linewidth}
		\includegraphics[width=1\linewidth]{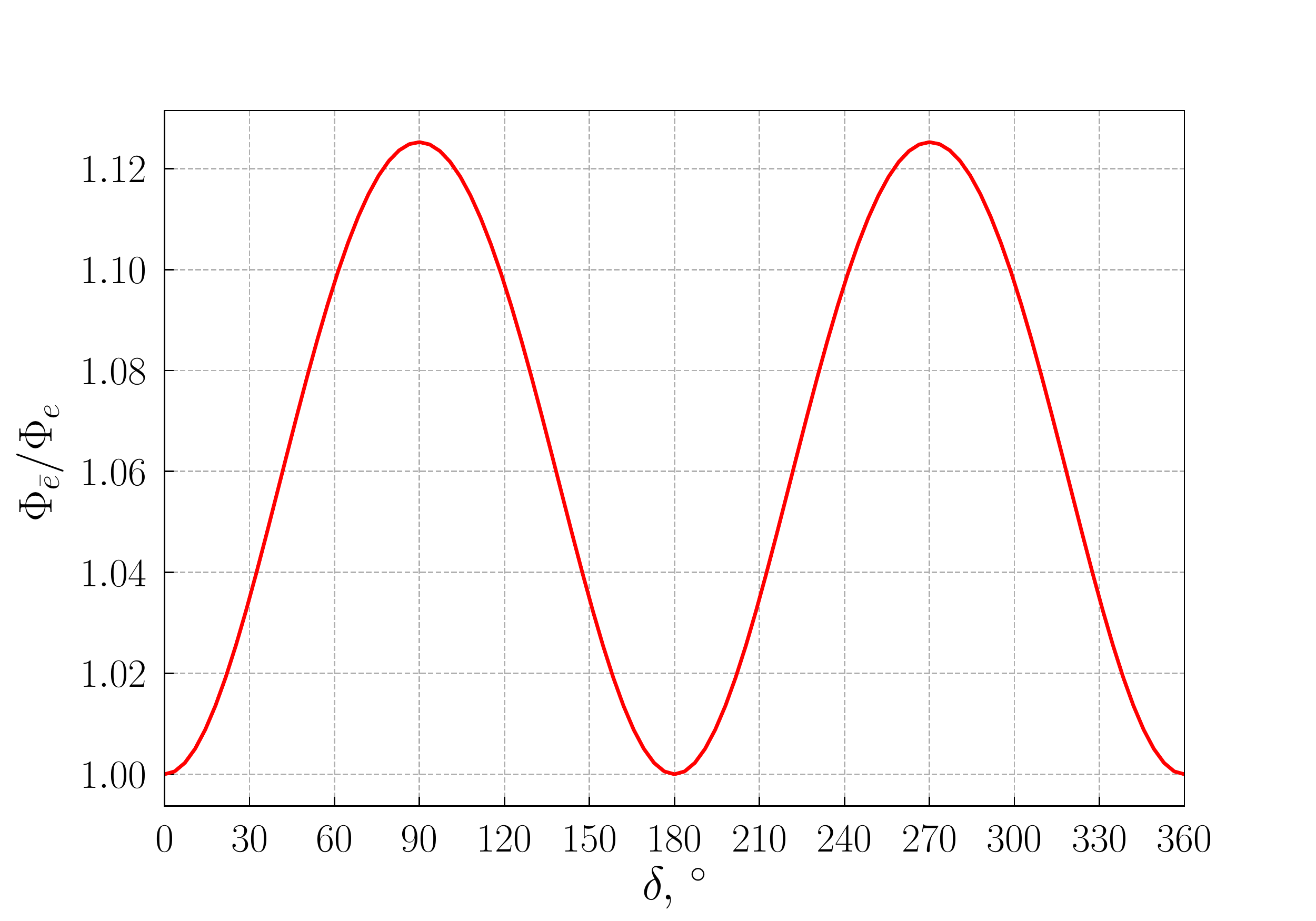}
	\end{minipage}
	\begin{minipage}{0.49\linewidth}
		\includegraphics[width=1\linewidth]{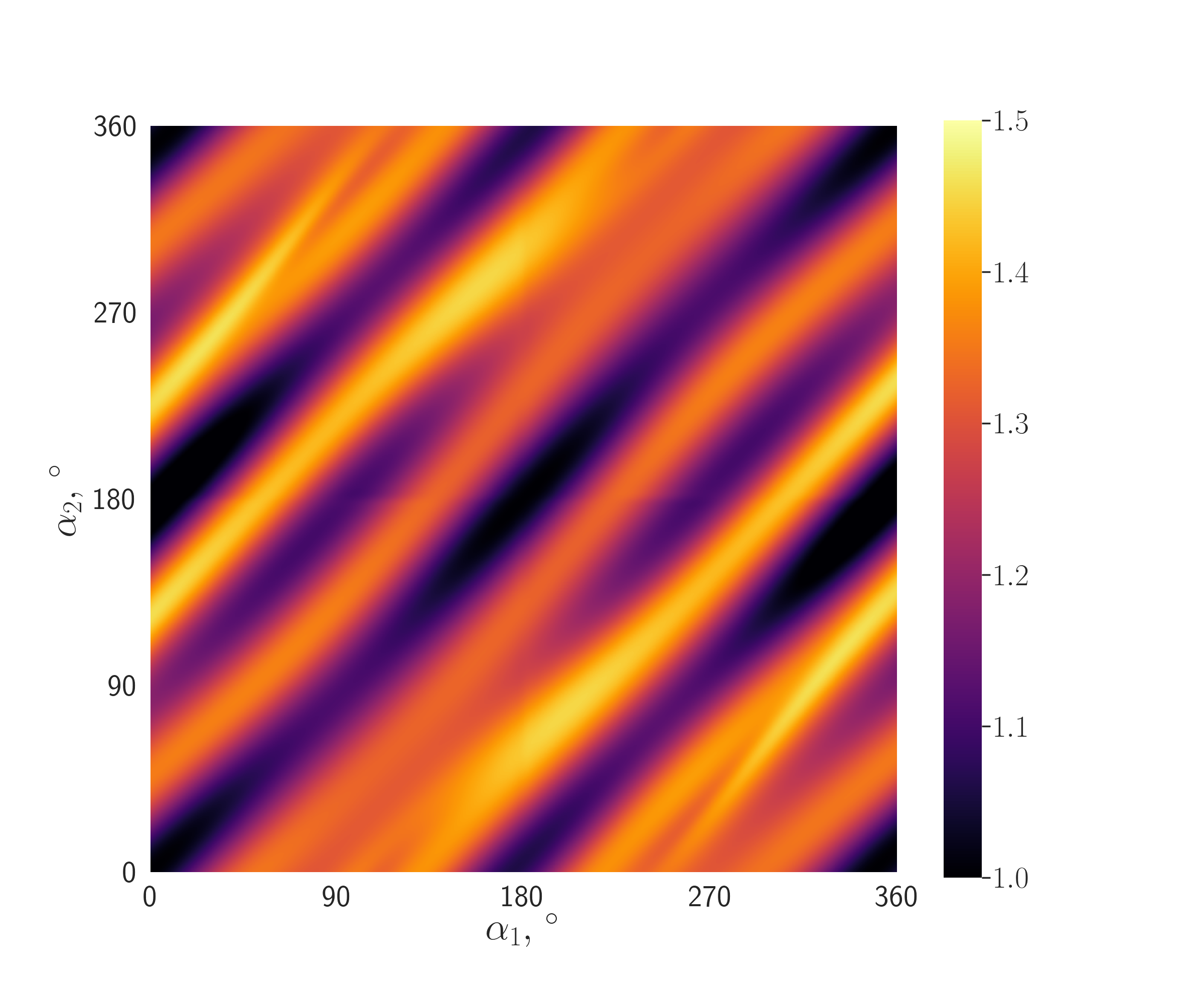}
	\end{minipage}
	\caption{$\bar{\nu}_e/\nu_e$ disproportion engendered by neutrino-antineutrino oscillations in a magnetic field. Left: as a function of the Dirac CP violating phase $\delta$. Right: as a function of the Majorana phases $\alpha_1$ and $\alpha_2$.}
\end{figure*}

\begin{figure*}
	\begin{minipage}{0.49\linewidth}
		\includegraphics[width=1\linewidth]{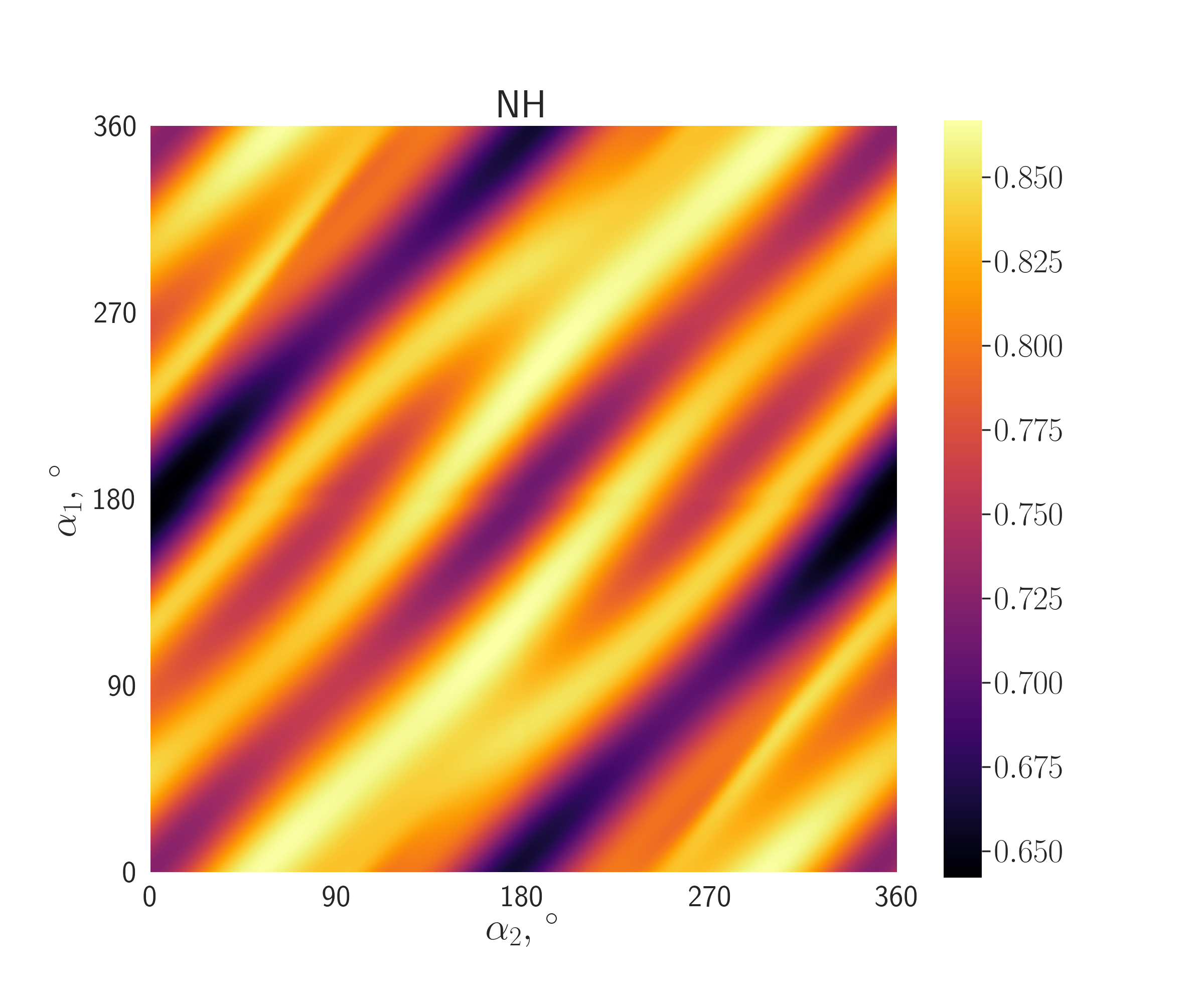}
	\end{minipage}
	\begin{minipage}{0.49\linewidth}
		\includegraphics[width=1\linewidth]{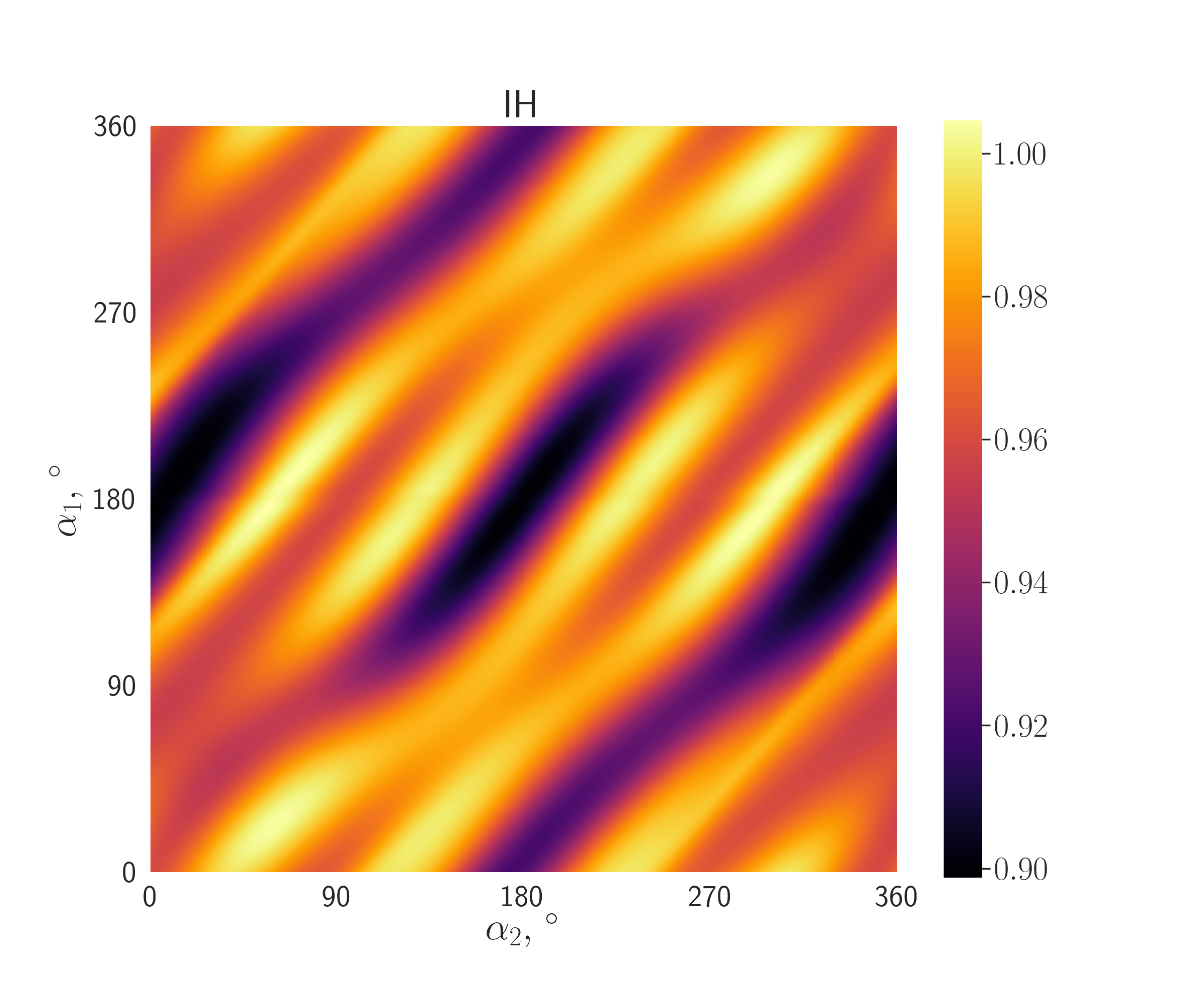}
	\end{minipage}
	\caption{$\bar{\nu}_e/\nu_e$ ratio outside supernova as a function of the Majorana CP violating phases $\alpha_1$ and $\alpha_2$ for NH and IH.}
\end{figure*}

\end{document}